\documentclass[reprint,amssymb, amsmath, aps, superscriptaddress, longbibliography, showpacs, footinbib, pra]{revtex4-1}
\usepackage{graphicx,epstopdf} 
\usepackage{amsmath}

\usepackage{color}
\usepackage[colorlinks,bookmarks=false,citecolor=darkblue,linkcolor=red,urlcolor=blue]{hyperref}

\definecolor{darkred}{rgb}{0.7,0.0,0.0}

\definecolor{darkblue}{rgb}{0,0.02,0.45}

\definecolor{darkgreen}{rgb}{0.02,0.45,0.0}

\definecolor{violet}{rgb}{0.8,0.2,0.6}

\begin{document}

\title{Singlet ground state in the alternating spin-$1/2$ chain compound NaVOAsO$_4$}

\author{U. Arjun}
\affiliation{School of Physics, Indian Institute of Science Education and Research Thiruvananthapuram-695551, India}
\author{K. M. Ranjith}
\author{B. Koo}
\author{J. Sichelschmidt}
\affiliation{Max Planck Institute for Chemical Physics of Solids, Nöthnitzer Str. 40, 01187 Dresden, Germany}
\author{Y. Skourski}
\affiliation{Dresden High Magnetic Field Laboratory, Helmholtz-Zentrum Dresden-Rossendorf, 01314 Dresden, Germany}
\author{M. Baenitz}
\affiliation{Max Planck Institute for Chemical Physics of Solids, Nöthnitzer Str. 40, 01187 Dresden, Germany}
\author{A. A. Tsirlin}
\affiliation{Experimental Physics VI, Center for Electronic Correlations and Magnetism, Institute of Physics, University of Augsburg, 86135 Augsburg, Germany}
\author{R. Nath}
\email{rnath@iisertvm.ac.in}
\affiliation{School of Physics, Indian Institute of Science Education and Research Thiruvananthapuram-695551, India}

\date{\today}

\begin{abstract}
We present the synthesis and a detailed investigation of structural and magnetic properties of polycrystalline NaVOAsO$_4$ by means of x-ray diffraction, magnetization, electron spin resonance (ESR), and $^{75}$As nuclear magnetic resonance (NMR) measurements as well as density-functional band structure calculations. Temperature-dependent magnetic susceptibility, ESR intensity, and NMR line shift could be described well using an alternating spin-$1/2$ chain model with the exchange coupling $J/k_{\rm B}\simeq 52$~K and an alternation parameter $\alpha \simeq 0.65$. From the high-field magnetic isotherm measured at $T=1.5 $~K, the critical field of the gap closing is found to be $ H_{\rm c}\simeq 16$~T, which corresponds to the zero-field spin gap of $\Delta_0/k_{\rm B}\simeq 21.4$~K. Both NMR shift and spin-lattice relaxation rate show an activated behavior at low temperatures, further confirming the singlet ground state. The spin chains do not coincide with the structural chains, whereas the couplings between the spin chains are frustrated. Because of a relatively small spin gap, NaVOAsO$_4$ is a promising compound for further experimental studies under high magnetic fields.
\end{abstract}
\maketitle

\section{\textbf{Introduction}}
\begin{figure*}
	\includegraphics[scale=0.85]{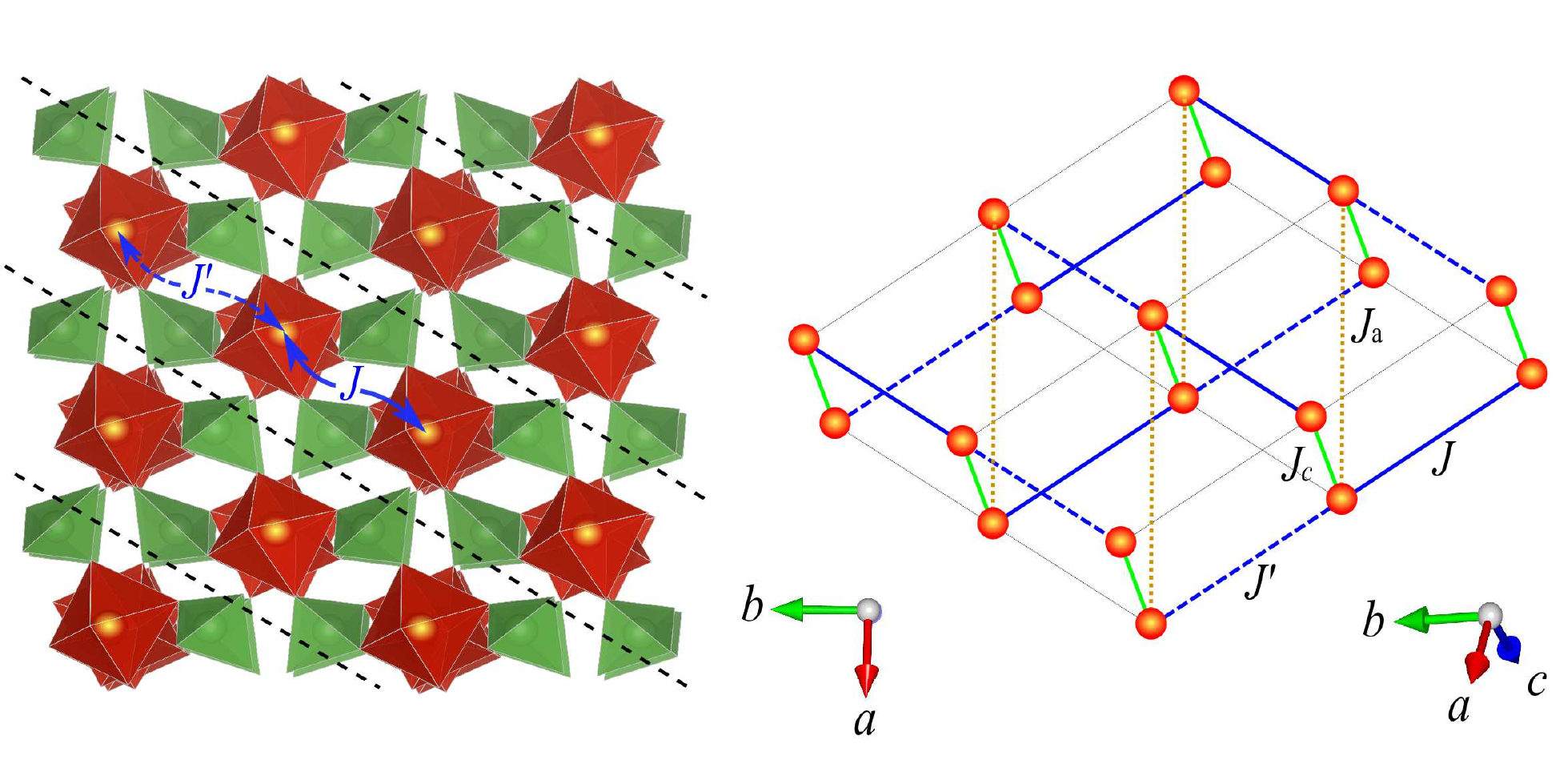}
	\caption{Left panel: crystal structure of NaVOAsO$_4$. The black dashed lines separate the spin chains. The VO$ _{6} $ and AsO$ _4 $ polyhedra are shown in red and green colors, respectively. Right panel: spin model showing the alternating ($J,J^\prime$) crossing chains and the frustrated inter-chain couplings in a different orientation.}
	\label{Fig1}
\end{figure*}

For decades, there has been a flourish of interest in one-dimensional (1D) quantum antiferromagnets (AFM), as they often provide unique opportunities to study the interplay between the charge, orbital, spin, and lattice degrees of freedom~\cite{Sachdev173}. In most of the 1D quantum AFMs, the frequently encountered ground state is a three dimensional (3D) long-range order (LRO), resulting from inter-chain interactions. Therefore, any new compound possessing a spin-gap in the excitation spectrum is of fundamental interest. Typically, the alternation of exchange interactions along the chain direction introduces a spin gap, as in (VO)$_2$P$_2$O$_7$~\cite{Yamauchi3729}. If this gap appears due to a structural distortion, which in turn causes spin dimerization, it is known as the spin-Peierls transition~\cite{Chesnut4677}. However, not every structural phase transition accompanied by a spin-gap formation is a spin-Peierls transition. Such a transition has only been observed in several organic compounds~\cite{Bray744,Huizinga4723} and CuGeO$_3$~\cite{Hase3651}. For the mixed-valence compound NaV$_2$O$_5$, the formation of a singlet ground state is reported to be due to charge ordering~\cite{Isobe1178}, while for NaTiSi$_2$O$_6$ it is due to orbital ordering~\cite{Isobe1423}.

Spin chains with the gap in the excitation spectrum are sensitive to perturbations like inter-dimer couplings and external magnetic field. Above a threshold limit, the spin gap is closed, and a multitude of field-induced phases emerge. The singlet state sustains in low fields, where the magnetization remains nearly zero. But a higher field can close the gap and trigger an AFM LRO above the critical field of the gap closing. This fascinating field-induced phenomenon can be described as the Bose-Einstein Condensation (BEC) of triplons, where the particle density is proportional to the applied magnetic field acting as chemical potential of the Bose gas~\cite{Rice760}. Notable compounds showing the triplon BEC are BaCuSi$_2$O$_6$, Sr$_3$Cr$_2$O$_8$, TlCuCl$_3$ etc~\cite{Sebastian617,Aczel207203,Ruegg62,Zapf2014,Freitas184426}.
Another important field-induced phenomenon shown by spin-gap systems is the emergence of a non-Fermi liquid-type Tomonaga-Luttinger liquid (TLL) phase above the critical field of the gap closing~\cite{Tomonaga544,Luttinger3}.
The spin-ladder compound (C$_{7}$H$_{10}$N)$_{2}$CuBr$_{4}$ presents an excellent experiment realization, where the TLL phase evolves from the singlet state~\cite{Jeong167206,Hong137207,Povarov2015,Jeong2016,Moeller2017}.
The transition between the 1D TLL and 3D BEC is successfully explained in another ladder compound (C$_{5}$H$_{12} $N)$_{2}$CuBr$_{4}$~\cite{Thielemann020408,Ruegg247202}.
Quasi-1D materials involving alternating spin chains may show signatures of both TLL and BEC physics, thus giving researchers an opportunity to study the crossover effects from 1D to 3D physics~\cite{Orignac140403,Willenberg060407}.

In addition, spin-gap materials also exhibit various other peculiar features like Wigner crystallization of magnons, magnetization plateaus, etc, under external magnetic field\cite{Kim017201}. Orthogonal dimer compound SrCu$_2$(BO$_3$)$_2$ with the Shastry-Sutherland lattice is a well-known example showing these exotic features~\cite{Kodama395}. The above field-induced effects are explored only in very few experimental systems till date due to the shortage of model compounds. In this context, materials with the small spin gap are preferable, so that one can explore the field-induced phenomena and the complete $H-T$ phase diagram using practically accessible magnetic fields.



In this work, we endeavor to elucidate the magnetic behavior of NaVOAsO$_{4}$ by means of magnetization, electron spin resonance (ESR), and $ ^{75} $As Nuclear Magnetic Resonance (NMR) experiments combined with density-functional band-structure calculations. A comparison is made with AgVOAsO$_{4}$ in order to understand the role of the inter-chain couplings on the size of the spin gap. AgVOAsO$_4$ is reported as the alternating spin-$1/2$ chain compound with the singlet ground state. It crystallizes in the monoclinic (space group $P2_1/c$) structure with the lattice parameters $a = 6.7157(1)$~\AA, $b = 8.8488(2)$~\AA, $c = 7.2848(2)$~\AA, $\beta = 115.28(1)^\circ$, and unit cell volume $V_{\rm cell} \simeq 391.5$~\AA$^3$~\cite{Tsirlin144412}. From the analysis of magnetic susceptibility, high-field magnetization, ESR, $ ^{75} $As NMR, and band-structure calculations, the dominant exchange coupling, alternation parameter, zero-field spin gap, and critical field of the gap closing are reported to be $ J/k_{\rm B} \simeq 41.8 $~K,  $\alpha = J^\prime/J \simeq 0.68 $, $\Delta_0/k_{\rm B} \simeq 13$~K, and $H_{\rm c}= \frac{\Delta}{g \mu_{\rm B}} \simeq 10$~T, respectively~\cite{Tsirlin144412,Ahmed224423}. 

When Ag$^+$ (ionic radius: 1.29~\AA) is replaced by Na$^+$ (ionic radius: 1.16~\AA), the crystal structure remains unaltered. The reported unit cell parameters for NaVOAsO$_4$ are $a = 6.650(1)$~\AA, $b=8.714(1)$~\AA, $c= 7.221(1)$~\AA, $\beta = 115.17(1)^\circ$, and $V_{\rm cell} \simeq 378.72$~\AA$ ^3 $~\cite{Haddad57}. 
The main objective of this work is to tune the spin gap and hence the critical field of the gap closing by chemical substitution, which may facilitate the study of field-induced effects using commonly accessible laboratory fields.

In NaVOAsO$_4$, there are one As atom, one Na atom, one V atom, and five nonequivalent O atoms. Similar to AgVOAsO$_4$, each AsO$ _4 $ tetrahedron is coupled with four VO$ _6 $ octahedra to form alternating spin chains that run nearly perpendicular to each other in the $ab$-plane, as shown in Fig.~\ref{Fig1}. There are also weak inter-chain exchange interactions, which make a frustrated network between the chains. The directions of all the exchange interactions are clearly depicted in the right panel of Fig.~\ref{Fig1}. The As atom is strongly coupled to the two neighbouring magnetic V$^{4+}$ ions along the direction of the spin chain. On the other hand, the Na atom is located in between the chains and seems to be very weakly coupled to the magnetic V$^{4+} $ ions. To the best of our knowledge, no information about the magnetic properties of this compound has been reported till date.

\section{Methods}
Polycrystalline sample of NaVOAsO$_{4}$ was synthesized by the conventional solid-state reaction technique by annealing the stoichiometric mixture of Na$_2$O (99.99\%), As$_2$O$_5$ (99.99\%), and VO$_2$ (99.99\%) (all from Sigma-Aldrich) in an evacuated silica tube at 500~$^\circ$C for 48~h with one intermediate grinding and pelletization. To avoid the hydration, the reactants were handled in an Ar filled glove box. The resulting sample was green in color. Its phase purity was confirmed by powder x-ray diffraction (XRD, PANalytical powder diffractometer with Cu $K_{\alpha}$ radiation, $\lambda_{\rm ave} = 1.54182$~\AA) at room temperature. Temperature ($T$) dependent powder XRD measurements were performed in the $T$-range 15~K~-~300~K using the low-$T$ attachment (Oxford Phenix) to the x-ray diffractometer. Rietveld refinement of the observed XRD patterns was performed using the \verb FullProf ~package~\cite{Carvajal55}, taking the initial parameters from Ref.~\onlinecite{Haddad57}.

DC magnetization ($M$) was measured as a function of $T$ and applied magnetic field $H$ using the vibrating sample magnetometer (VSM) attachment to the Physical Property Measurement System [PPMS, Quantum Design]. High-field magnetization isotherm ($M$ vs $H$) was measured at $T = 1.5$~K in pulsed magnetic field up to 60~T at the Dresden High Magnetic Field Laboratory~\cite{Tsirlin132407}.

The ESR experiments were carried out on a fine powdered sample with the standard continuous-wave spectrometer between 3~K and 300~K. We measured the power $P$ absorbed by the sample from a transverse magnetic microwave field (X-band, $\nu\simeq 9.4$~GHz), as a function of the external magnetic field $B$. The lock-in technique was used to improve the signal-to-noise ratio which yields the derivative of the resonance signal $dP/dB$. The $g$-factor was determined by the resonance condition $g=\frac{h\nu}{\mu_{\rm B} B_{\rm res}}$, where $h$ is the Planck's constant, $\mu_{\rm B}$ is the Bohr magneton, $ \nu $ is the resonance frequency, and $ B_{\rm res} $ is the corresponding resonance field.

The NMR experiments on $^{75}$As nucleus (nuclear spin $I = 3/2$, gyromagnetic ratio $\gamma/2\pi = 7.291$~MHz/T) were carried out using pulsed NMR technique at a radio frequency $ \nu $ ($= \gamma $/2$ \pi H$) of 50.44~MHz, which corresponds to the magnetic field of $ H\approx6.8 $~T. At each temperature, the $^{75}$As NMR spectrum was obtained by sweeping the magnetic field while keeping the frequency fixed. The NMR shift $K(T) = [H_{\rm ref}-H(T)]/H(T)$ was determined from the resonance field $ H(T) $ of the sample with respect to the resonance field $ H_{\rm ref}$ of the reference GaAs sample. The spin-lattice relaxation rate ($1/T_1$) measurements were done using the standard inversion recovery method.

Exchange couplings of the Heisenberg spin Hamiltonian
\begin{equation}
 \hat H=\sum_{\langle ij\rangle}J_{ij}\mathbf S_i\mathbf S_j,
\end{equation}
where $S=\frac12$ and the summation is over the lattice bonds $\langle ij\rangle$, were obtained by a mapping procedure using density-functional (DFT) band-structure calculations performed in the \texttt{FPLO} code~\cite{fplo} within the generalized gradient approximation (GGA) for the exchange-correlation potential~\cite{pbe96}. The mean-field DFT+$U$ correction for correlation effects in the V $3d$ shell was used. Additionally, we analyzed the uncorrelated GGA band structure and extracted the hopping parameters, which were further used to obtain antiferromagnetic exchange couplings on the level of second-order perturbation theory, as further explained in Sec.~\ref{sec:dft}. The typical $k$-mesh included 128 points in the irreducible part of the first Brillouin zone. In the DFT+$U$ calculations, we applied $U_d=4$\,eV and $J_d=1$\,eV for the on-site Coulomb repulsion and Hund's exchange, respectively, as well as the around-mean-field correction for double-counting~\cite{Tsirlin144412,Nath024418,Tsirlin2011c}. The structural parameters from Ref.~\onlinecite{Haddad57} were used.

Magnetization curves were simulated using the \verb|dirloop_sse|~\cite{dirloop} algorithm of the \texttt{ALPS} simulation package~\cite{alps} on two-dimensional finite lattices with periodic boundary conditions and up to 256 sites.

\section{Results and Discussion}
\subsection{X-Ray diffraction}
\begin{figure}
	\includegraphics[scale=1.1]{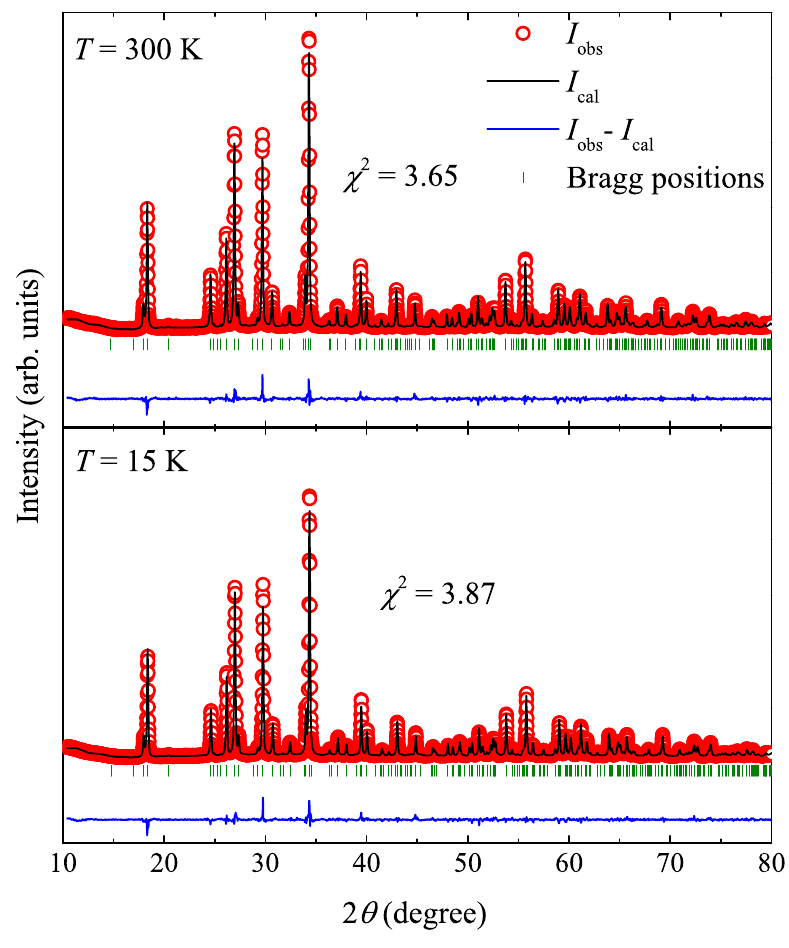}
	\caption{(Color online) Powder x-ray diffraction pattern (open circles) for NaVOAsO$_{4}$ at two different temperatures ($T=15$~K and 300~K). The solid line represents the full-profile refinement, with the vertical bars showing the expected Bragg peak positions, and the lower solid blue line representing the difference between the observed and calculated intensities.}
	\label{ref}
\end{figure}

Figure~\ref{ref} shows the powder XRD pattern of NaVOAsO$_{4}$ at two different temperatures, $T=15$~K and 300~K, along with the refinement. The obtained best fit parameters at 300~K and 15~K are [$a = 6.6469(1)$~\AA, $b = 8.7086(1)$~\AA, $c = 7.2156(1)$~\AA, $\beta = 115.21(1)^\circ$, $V_{\rm cell} \simeq 377.89$~\AA$^3$, and the goodness-of-fit $\chi^2 \simeq 3.65$] and [$a = 6.6343(1)$~\AA, $b = 8.6952(2)$~\AA, $c = 7.2034(2)$~\AA, $\beta = 115.177(1)^\circ$, $V_{\rm cell} \simeq 376.06$~\AA$^3$, and $\chi^2 \simeq 3.87$], respectively. These values are in close agreement with the reported values~\cite{Haddad57}.
A slight decrease in volume at room temperature for NaVOAsO$_4$ compared to AgVOAsO$_4$ is likely due to the fact that Na$^+$ (ionic radius: 1.16~\AA) has a smaller ionic radius than Ag$^+$ (ionic radius: 1.29~\AA).
The atomic positions obtained from the structural refinement at room temperature ($T = 300$~K) are listed in Table~\ref{refinement}. 

\begin{figure}
	\includegraphics[scale=1.1]{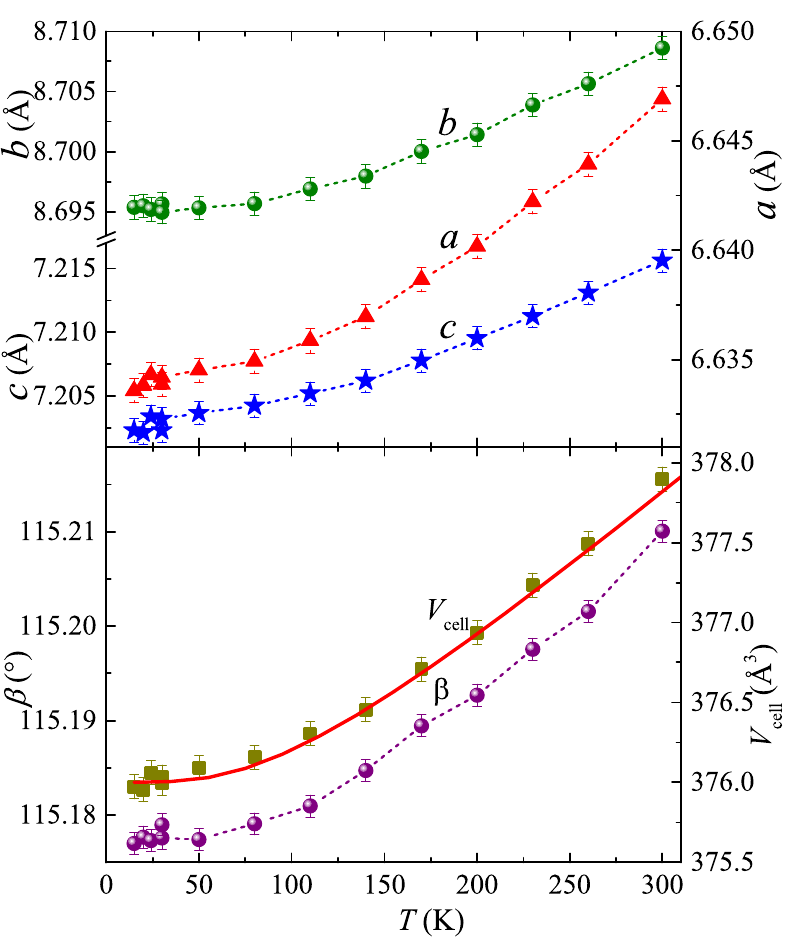}
	\caption{The lattice constants ($a$, $b$, $c$), monoclinic angle ($\beta$), and unit cell volume ($V_{\rm cell}$) are plotted as a function of temperature. The solid line in the bottom panel represents the fit using Eq.~\eqref{Vcell}.}
	\label{Fig3}
\end{figure}

The powder XRD pattern at different temperatures were also analyzed by the Rietveld refinement to understand the temperature variation of the crystal structure. Going from 300~K to 15~K, the overall XRD pattern remains intact, thus ruling out any structural phase transitions or lattice distortions that take place in other spin-gap compounds like CuGeO$_3$~\cite{Hirota736}, NaV$_2$O$_5$~\cite{Fuji326}, and NaTiSi$_2$O$_6$~\cite{Isobe1423}. The temperature dependence of the lattice parameters obtained from the refinement is plotted in Fig.~\ref{Fig3}. The lattice constants $a$, $b$, and $c$, and the angle $\beta$ are found to decrease upon cooling. As a result, $V_{\rm cell}$ also shows a gradual decrease with decreasing temperature from 300~K to 15~K.

\begin{table}
	\caption{Atomic coordinates and isotropic atomic displacement parameters ($U_{\rm iso}$, in units of $10^{-2}$~\AA$^2$) of NaVOAsO$_{4}$ obtained from the Rietveld refinement of the powder XRD data at room temperature (300~K). The crystal structure is monoclinic with the space group $P2_1/c$ (No. 14). The obtained lattice parameters are $a = 6.6469(1)$~\AA, $b = 8.7086(1)$~\AA, $c = 7.2156(1)$~\AA, $\beta = 115.21(1)^\circ$, and $V_{\rm cell} \simeq 377.89$~\AA$^3$. The $U_{\rm iso}$ values for oxygen were constrained. The error bars are from the Rietveld refinement.}
	\label{refinement}
	\begin{tabular}{ccccccc}
		\hline \hline
		Atom & Wyckoff & $x$ & $y$ & $z$ & $U_{\rm iso}$\\
		&position&&&&\\\hline
		Na & $4e$ & 0.261(1) & 0.5840(4) & 0.285(1) &0.16(4)\\
		V & $4e$ & 0.2414(7) & 0.2610(5) & 0.0299(6) &1.0(1)\\
		As & $4e$ & 0.7483(7) & 0.4366(1) & 0.2492(6) &0.93(4)\\
		O1 & $4e$ & 0.251(3) & 0.3217(7) & 0.261(2) &0.48(5)\\ 
		O2 & $4e$ & 0.081(2) & 0.801(1) & 0.090(1) &0.48(5)\\
		O3 & $4e$ & 0.125(2) & 0.037(1) & 0.382(2) &0.48(5)\\
		O4 & $4e$ & 0.542(2) & 0.319(1) & 0.078(1) &0.48(5)\\
		O5 & $4e$ & 0.363(2) & 0.045(1) & 0.116(2) &0.48(5)\\
		\hline \hline
	\end{tabular}
\end{table}
The temperature variation of $V_{\rm cell}$ was modeled following the Gr\"{u}neisen approximation for the zero-pressure state, where the effects of thermal expansion are considered to be equivalent to the elastic strain~\cite{Wallace1998}. $ V_{\rm cell}(T) $ can be written as~\cite{Pallab2018}
\begin{equation}\label{Vcell}
V_{\rm cell}(T)= \gamma U(T)/K_0 + V_0,
\end{equation}
where $ V_0 $ is the unit cell volume at $T=0$ K, $K_0$ is the bulk modulus, $\gamma$ is the Gr\"{u}neisen parameter, and $ U(T) $ is the internal energy. The $ U(T) $ can be expressed in terms of the Debye approximation as
\begin{equation}\label{U}
U(T) = 9Nk_{\rm B}T\left(\frac{T}{\theta_{\rm D}}\right)^3\int_0^{\frac{\theta_{\rm D}}{T}}\frac{x^3}{(e^x-1)} dx,
\end{equation}
where $\theta_{\rm D}$ is the characteristic Debye temperature, $N$ is the number of atoms per unit cell, and $k_{\rm B}$ is the Boltzmann constant. The fit is shown as the solid line in the bottom panel of Fig.~\ref{Fig3}. The obtained best fit parameters are $\theta_{\rm D}\simeq 375$~K and $ V_0 \simeq 376$~\AA$^3$.

\subsection{Magnetization}
\begin{figure}
		\includegraphics[scale=1.1]{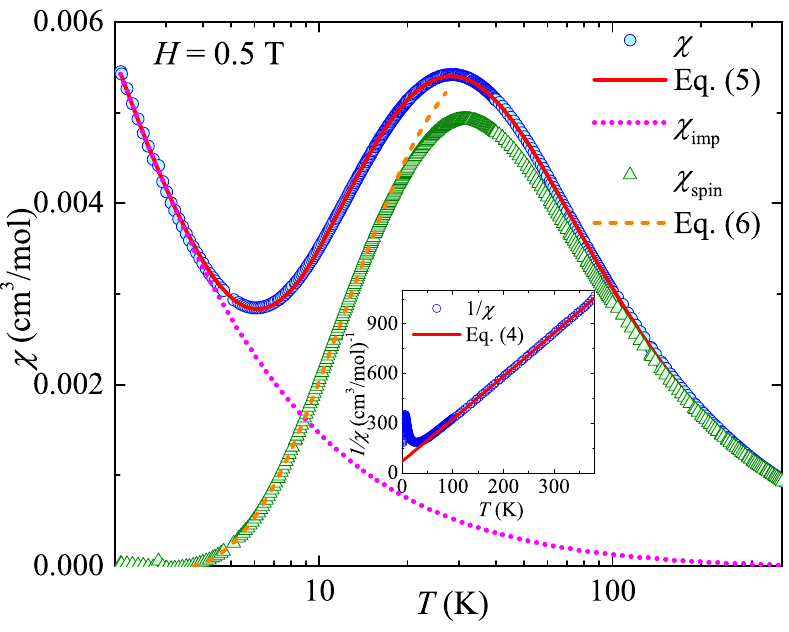}
		\caption{$\chi(T)$ measured at $H = 0.5$~T with the solid line representing the fit using Eq.~\eqref{chi_alt}. The dotted line shows the $ \chi_{\rm imp} $, and the dashed line shows the fit on the $ \chi_{\rm spin} $ data using Eq.~\eqref{Dample}. Inset: $1/\chi(T)$ data along with the fit using Eq.~\eqref{cw} (solid line).}
		\label{Fig4}
\end{figure}

The temperature-dependent magnetic susceptibility $\chi(T)$ ($ \equiv M/H $) measured in an applied field of $H = 0.5$~T is shown in Fig.~\ref{Fig4}. As the temperature decreases, $\chi(T)$ increases in the Curie-Weiss manner as expected in the paramagnetic regime and then shows a broad maximum at around $T_\chi^ {\rm max} \simeq 29$~K, indicative of a short-range magnetic order, as expected for low-dimensional spin systems. Below $T_\chi^ {\rm max}$, it shows a rapid decrease and then an upturn. This low temperature upturn is likely due to the extrinsic paramagnetic impurities and/or defects present in the sample~\cite{Arjun174421}. There is no clear indication of any magnetic LRO down to 2~K.

To extract the magnetic parameters, $\chi(T)$ at high temperatures was fitted by the following expression
\begin{equation}\label{cw}
\chi(T) = \chi_0 + \frac{C}{T - \theta_{\rm CW}},
\end{equation}
where $\chi_0$ is the temperature independent contribution consisting of the diamagnetic susceptibility ($\chi_{\rm core}$) of core electron shells and Van-Vleck paramagnetic susceptibility ($\chi_{\rm VV}$) of the open shells of the V$^{4+}$ ions. The second term in Eq.~\eqref{cw} is the Curie-Weiss (CW) law with the CW temperature $\theta_{\rm CW}$ and Curie constant $C = N_{\rm A} \mu_{\rm eff}^2/3k_{\rm B}$. Here, $N_{\rm A}$ is Avogadro's number, $\mu_{\rm eff} = g\sqrt{S(S+1)}$$\mu_{\rm B}$ is the effective magnetic moment, $g$ is the Land$\acute{\rm e}$ $g$-factor, $\mu_{\rm B}$ is the Bohr magneton, and $S$ is the spin quantum number. Our fit in the temperature range 150~K to 380~K (inset of Fig.~\ref{Fig4}) yields $\chi_0 \simeq -2.46 \times 10^{-4}$~cm$^3$/mol, $C \simeq 0.39$~cm$^3$K/mol, and $\theta_{\rm CW} \simeq -27.4$~K.
From the value of $C$, the effective moment was calculated to be $\mu_{\rm eff} \simeq 1.76 \mu_{\rm B}$ in good agreement with the expected spin-only value of 1.73~$\mu_{\rm B}$ for $S = 1/2$, assuming $g = 2$. The negative value of $\theta_{\rm CW}$ indicates that the dominant exchange couplings between V$^{4+}$ ions are antiferromagnetic in nature.

\begin{figure}
	\includegraphics[scale=1.1]{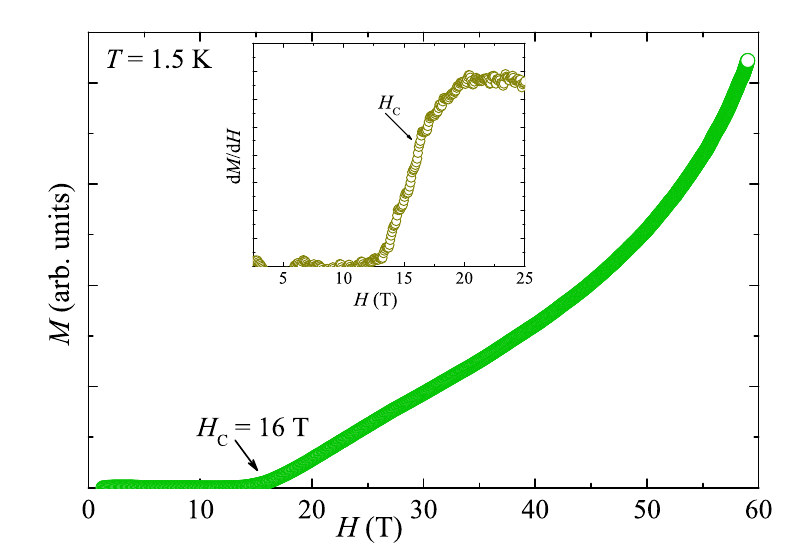}
	\caption{Magnetization ($M$) vs $H$ measured at $ T=1.5 $ K (after subtracting the paramagnetic background) using pulsed magnetic field. The critical field of gap closing $ H_{\rm c} \simeq 16$~T is marked with an arrow. Inset: $dM/dH$ vs $H$ highlights the position of $H_{\rm c}$.}  
	\label{Fig5}
\end{figure}

In order to estimate the exchange couplings, the $\chi(T)$ data were fitted by 
\begin{equation}\label{chi_alt}
\chi(T) = \chi_0 + \frac {C_{\rm imp}}{T-\theta_{\rm imp}}+\chi_{\rm spin}.
\end{equation}
Here, $C_{\rm imp}$ represents the Curie constant corresponding to the impurity spins and $\theta_{\rm imp}$ quantifies the effective interaction between them. $\chi_{\rm spin}$ is the expression for the spin susceptibility of the alternating spin-$1/2$ chain model, valid for the entire range $0 \leq \alpha~(=J ^\prime /J) \leq 1$ and for the entire temperature range $k_{\rm B}T/J\geq0.01$, where $J ^\prime$ and $J$ are the exchange couplings along the chain.\cite{Johnston9558} As shown in Fig.~\ref{Fig4}, Eq.~\eqref{chi_alt} fits very well to the $\chi(T)$ data over the whole temperature range, yielding $J/k_{\rm B} \simeq 51.6$~K, $\alpha \simeq 0.65$, $\chi_0 \simeq -3.32 \times 10^{-5}$~cm$^3$/mol, $C_{\rm imp} \simeq 0.01626$~cm$^3$K/mol, and $\theta_{\rm imp} \simeq -0.9$~K. During the fitting procedure, the value of $g$ was fixed to $ g \simeq 1.99 $, obtained from the ESR experiments (discussed later). The above value of $C_{\rm imp}$ corresponds to the concentration of nearly 4.1~\%, assuming the impurity spins to be $S = 1/2$. Using the value of $J/k_{\rm B}$ and $\alpha$, the spin gap is estimated to be $\Delta_0/k_{\rm B} = J/k_{\rm B}[(1 - \alpha)^{3/4}(1 + \alpha)^{1/4} \simeq 26.5$ K~\cite{Johnston9558}. The critical field for closing a gap of $\Delta_0/k_{\rm B} \simeq 26.5$~K is estimated to be $H_{\rm c} = \Delta_0/(g \mu_{\rm B}) \simeq 19.8$~T.

In order to see the low-temperature gapped behavior, the impurity contribution ($\chi_{\rm imp} =\chi_0 + \frac {C_{\rm imp}}{T-\theta_{\rm imp}}$) was subtracted from $\chi(T)$ and the obtained $\chi_{\rm spin}(T)$ is plotted in the same Fig.~\ref{Fig4}. It shows a clear exponential decrease towards zero demonstrating the singlet ground state. $\chi_{\rm spin}(T)$ below the broad maximum is fitted using the expression for the spin susceptibility of a gapped spin-$1/2$ 1D Heisenberg chain~\cite{Sachdev943}
\begin{equation}\label{Dample}
\chi_{\rm 1D}(T) \propto \sqrt{\frac{\Delta^{\chi}_{\rm 1D}}{k_{\rm B}T}}\,e^{-{\Delta^{\chi}_{\rm 1D}}/k_{\rm B}T}.
\end{equation}
The fit in the low-$T$ region ($T\leq 15$~K) returns $ \Delta^{\chi}_{\rm 1D}/k_{\rm B}\simeq 22.4 $~K which is slightly lower than the value obtained from the alternating chain model fit. The critical field for closing this gap is $H_{\rm c}\simeq 16.7$~T.

The magnetic isotherm measured up to 60~T at $T=1.5$~K is shown in Fig.~\ref{Fig5}. In the low-field regime, $M$ is very small and remains almost constant up to about 16~T, suggesting that the critical field of the gap closing is $H_{\rm c} \simeq 16$~T. It is slightly higher than $H_c\simeq 10$\,T in AgVOAsO$_4$ but is still in the accessible range. This value of $H_{\rm c}$ corresponds to the zero-field spin gap of $ \Delta_0/k_{\rm B} \simeq 21.4$~K, which is slightly lower than the one obtained from the $ \chi(T) $ analysis but consistent with that obtained from the fit using Eq.~\eqref{Dample}. Typically, in the spin-gap compounds, the magnetization remains zero up to $ H_{\rm c} $. However, in our compound, a finite value of the moment was obtained below $H_{\rm c}$ which is likely due to the saturation of paramagnetic impurities in the powder sample. The data shown in Fig.~\ref{Fig5} are corrected for this paramagnetc contributions. For $ H \geq 16 $~T, $M$ increases almost linearly with $H$ and then shows a pronounced curvature. It does not saturate even at 60~T. A pronounced curvature above $H_{\rm c}$ reflects strong quantum fluctuations, as expected for 1D spin chains. The lack of saturation even at 60~T suggests a large value of the exchange coupling.

\subsection{ESR}
\begin{figure}
	\includegraphics[scale=1.1]{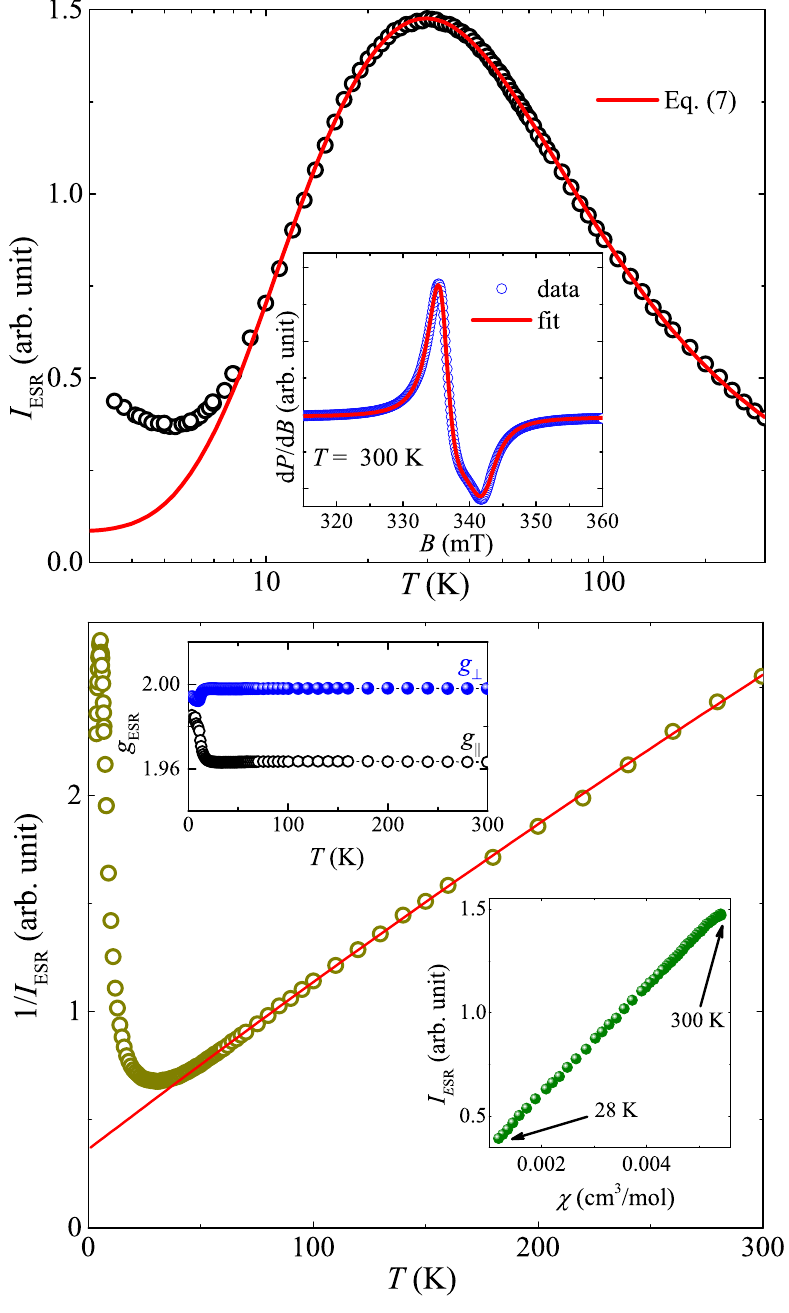}
	\caption{Upper panel: Temperature-dependent ESR intensity, $I_{\rm ESR}(T)$, obtained by integrating the ESR spectra of the polycrystalline NaVOAsO$_4$ sample. The solid line represents the fit as described in the text. Inset: A typical spectrum (symbols) at $ T= 300 $ K together with the fit using the powder-averaged Lorentzian shape for the uniaxial $g$-factor anisotropy. Lower panel: $1/I_{\rm ESR}$ vs $T$ along with the fit (see the text). Top inset: Temperature-dependent $ g $-factor ($g_{\parallel}$ and $g_{\perp})$ obtained from the Lorentzian fit. Bottom inset: $I_{\rm ESR}$ vs $\chi$ with temperature as an implicit parameter.}
	\label{Fig6}
\end{figure}
ESR results on the NaVOAsO$_4$ powder sample are presented in Fig.~\ref{Fig6}. The inset of the upper panel of Fig.~\ref{Fig6} depicts a typical ESR powder spectrum at room temperature. We tried to fit the spectra using the powder-averaged Lorentzian line for the uniaxial $g$-factor anisotropy. The fit reproduces the spectral shape very well at $T=300$~K, yielding the anisotropic $g$-tensor components $g_{\parallel} \simeq 1.963$ (parallel) and $g_{\perp} \simeq 1.998$ (perpendicular). The isotropic $g \left[ =\sqrt{(g^{2}_{\parallel}+2g^{2}_{\perp})/3}\right]$ value is calculated to be $\sim 1.99$. Such a value of $g$ is typically observed for V$^{4+}$ compounds~\cite{Yogi024413,Tsirlin144412}. As shown in the top inset of the lower panel of Fig.~\ref{Fig6}, both $g_{\parallel}$ and $g_{\perp}$ are found to be temperature-independent down to about 10~K. Only below 10~K, an anomalous behaviour was observed, which could be due to extrinsic foreign phases. The ESR intensity ($ I_{\rm ESR} $) as a function of temperature shows a pronounced broad maximum at $ \sim 30$~K, similar to the bulk $ \chi(T) $ data. In the bottom inset of the lower panel of Fig.~\ref{Fig6}, $ I_{\rm ESR} $ is plotted as a function of $\chi$ with temperature as an implicit parameter. It shows linearity over the temperature range of 28~K to 300~K, implying that $ I_{\rm ESR} $ traces $ \chi(T) $ nicely.

In order to estimate the exchange couplings, $ I_{\rm ESR}(T)$ data were fitted by the equation
\begin{equation}\label{ESR_alt}
I_{\rm ESR} = A + B\chi_{\rm spin}(T),
\end{equation}
where $A$ and $B$ are arbitrary constants. The fit in the range 9~K to 300~K (upper panel of Fig.~\ref{Fig6}) with the fixed $g \simeq 1.99$ yields $J/k_{\rm B} \simeq 49.0$~K and $\alpha \simeq 0.64$. This value of $J/k_{\rm B}$ is close to the one obtained from the $\chi(T)$ analysis. In the lower panel of Fig.~\ref{Fig6}, the inverse ESR intensity ( $1/I_{\rm ESR}$) is plotted as a function of temperature. At high temperatures, the linear behavior resembles the $ \chi^{-1} $ data. We fitted $1/I_{\rm ESR}(T)$ by an expression
\begin{equation}\label{ESR_inverse}
1/I_{\rm ESR} =\left[ M + \frac{N}{T-\theta_{\rm CW}}\right]^{-1},
\end{equation}
where $M$ and $N$ are arbitrary constants. The fit in the high-temperature region yields $ \theta_{\rm CW} \simeq -45$~K, which is little higher in magnitude than the value obtained from the $ \chi^{-1} $ analysis.

\subsection{$^{75}$As NMR}
NMR is a powerful local technique to study the static and dynamic properties of a spin system. In NaVOAsO$_{4}$, the $^{75}$As nuclei are strongly hyperfine-coupled to the magnetic V$^{4+}$ ions along the spin chains. Therefore, the low-lying excitations of the V$^{4+}$ spins can be probed by means of the $^{75}$As NMR spectra, NMR shift, and spin-lattice relaxation time measurements. As discussed earlier, the $\chi(T)$ data do not show an exponential decrease as anticipated for a spin gap system since the low-temperature impurity contribution masks the spin-gap behavior. Secondly, due to several fitting parameters, the $ \chi(T) $ analysis often doesn't provide reliable magnetic parameters. In this context, the NMR shift has an advantage over the bulk $\chi(T)$. The bulk $\chi(T)$ data include additional contributions coming from impurities and defects present in the sample, whereas the NMR shift directly measures the intrinsic spin susceptibility and is completely free from the extrinsic contributions. Therefore, one can precisely estimate the magnetic parameters by analyzing the temperature-dependent NMR shift instead of the bulk $\chi(T)$, and underpin the singlet ground state.

\subsubsection{NMR Spectra}
$^{75}$As is a quadrupole nucleus with the nuclear spin $ I = 3/2 $ in a non-cubic environment. Therefore, the four-fold degeneracy of the nuclear spin $ I = 3/2 $ is partially lifted by the interaction between the nuclear quadrupole moment $ Q $ and the surrounding electric-field gradient (EFG). The nuclear spin Hamiltonian can be written as a sum of the Zeeman and quadrupolar interactions~\cite{Curro026502,Slichter1992},
\begin{equation}
\mathcal{H} = -\gamma \hbar {\hat I_{\rm z}} H(1+K) + \frac{h\nu_Q}{6}[(3\hat{I}_{z^\prime}^2-\hat{I}^2)+\eta(\hat{I}_{x^\prime}^2-\hat{I}_{y^\prime}^2)].
\end{equation}
Here, the nuclear quadrupole resonance (NQR) frequency is defined as $ \nu_{\rm Q}=  \nu_{\rm z}\sqrt{1+\eta^2/3}=\frac{3e^2qQ}{2I(2I-1)h} \sqrt{1+\eta^2/3}$, $e$ is the electron charge, $\hbar~(=h/2\pi)$ is the Planck's constant, $H$ is the applied field along $\hat{z}$, $K$ is the magnetic shift due to hyperfine field at the nuclear site, $ V_{\alpha \beta} $ are the components of the EFG tensor, $ eq = V_{z^\prime z^\prime} $ is the largest eigenvalue of the EFG, and $ \eta $ ($=\mid V_{x^\prime x^\prime}-V_{y^\prime y^\prime}\mid/V_{z^\prime z^\prime}$) is the asymmetry parameter. The principal axes $\{ x^\prime $, $ y^\prime $, $ z^\prime \}$ of the EFG tensor are defined by the local symmetry of the crystal structure. Therefore, resonance frequency corresponding to any nuclear transition is strongly dependent on the direction of the applied field relative to the crystallographic axes. The parameters $ \nu_{\rm z} $, $\eta$, and $\hat{z}^\prime$ can fully characterize the EFG, where, $\hat{z}^\prime$ is the unit vector in the direction of the principal axis of the EFG with the largest eigenvalue.

When the Zeeman term dominates over the quadrupole term, first-order perturbation theory is enough for describing the system. In such a scenario, for a $ I=3/2 $ nucleus, two equally spaced satellite peaks ($I_{z} = \pm 3/2 \longleftrightarrow \pm 1/2$) should appear on either side of the central peak ($I_{z} = +1/2 \longleftrightarrow -1/2$), separated by $ \nu_Q $. On the other hand, when the quadrupole effects are large enough, second-order perturbation theory is required, and and the peak positions depend strongly on the angle $\theta$ between the applied field $H$ (along $ \hat{z} $) and $\hat{z}^\prime$.
The expression for the resonance frequency for the central transition can be written as
\begin{eqnarray}\label{peak}
&&\nu_{(\pm \frac{1}{2})}=\nu_0+\frac{\nu_Q^2}{32\nu_0}(1-{\rm cos}^2\theta) \nonumber\\ 
&&\left[ (54{\rm cos}^2\theta)
\left(1+\frac{2}{3}\eta {\rm cos}2\phi\right) -6 \left(1-\frac{2}{3}\eta {\rm cos}2\phi\right)\right]  \nonumber\\ &&+\frac{\eta^2\nu_Q^2}{72\nu_0}\left[-12-18{\rm cos}^2\theta - \frac{27}{2}{\rm cos}^2 2\phi({\rm cos}^2\theta-1)^2 \right]
\end{eqnarray}
where $ \nu_0 $ is the Larmor frequency and $\phi$ is the azimuthal angle. For a polycrystalline sample, the NMR spectra are typically broad due to random distribution of the internal field. They use to show the central transition ($I_{z} = +1/2 \longleftrightarrow -1/2$) split into two horns, which correspond to crystallites with $ \theta\simeq 41.8 ^\circ$ (lower frequency peak) and to $ \theta=90 ^\circ$ (upper frequency peak)~\cite{Slichter1992,Grafe047003}.

The $ ^{75} $As NMR spectra measured at different temperatures are shown in Fig.~\ref{Fig7}. Their asymmetric double-horn line shape can be described well by the second-order nuclear quadrupolar interaction. The obtained fitting parameters corresponding to the spectrum at $T=115$~K are: the isotropic shift $ K_{\rm iso}\simeq 1.8$~\%, axial shift $K_{\rm axial}\simeq -0.031$~\%, anisotropic shift $ K_{\rm aniso}\simeq 0$~\%, NQR frequency $ \nu_Q\simeq 6.12$~MHz, asymmetry parameter $ \eta\simeq 0.589 $, and line width $\simeq 48.7$ MHz. The quadrupolar frequency is found to be almost constant in the measured temperature range, thus ruling out the possibility of any structural deformation of VO$ _6 $ octahedra and/or lattice distortion. This is in sharp contrast to what is observed in the spin-Peierls compound CuGeO$_3$ where the lattice distortion leads to the spin dimerization~\cite{Hase3651} and in $\alpha^\prime-$Na$_x$V$_2$O$_5$ where the singlet ground state is driven by charge ordering~\cite{Revurat4176} at low temperatures. This is indeed consistent with our temperature dependent powder XRD data.

\subsubsection{NMR Shift}
\begin{figure}
	\includegraphics[scale=1.0]{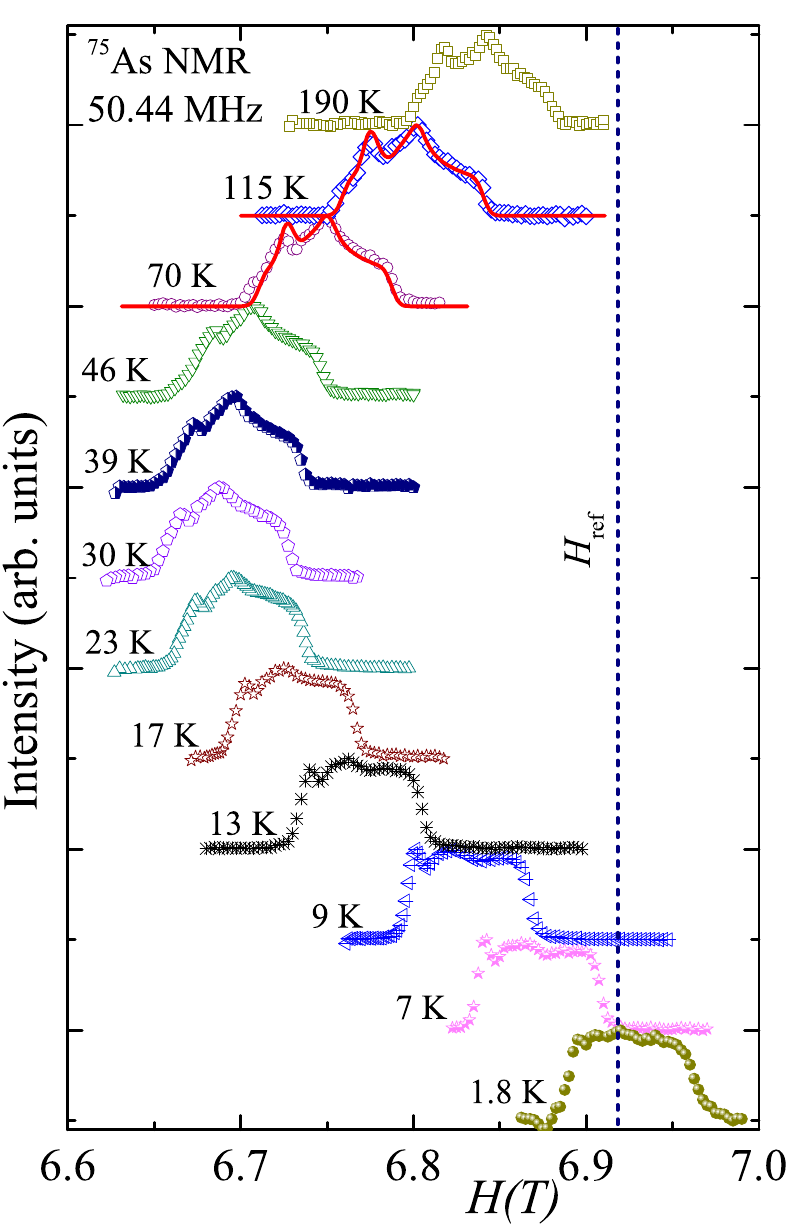}
	\caption{Temperature-dependent field-sweep $^{75}$As NMR spectra of NaVOAs$_4$ measured at $\nu = 50.44$~MHz. The solid lines are the fits to the spectra at $T = 70$~K and 115~K. The vertical dashed line corresponds to the $^{75}$As resonance position of the reference GaAs sample.}
	\label{Fig7}
\end{figure}

\begin{figure}
	\includegraphics[scale=1.1]{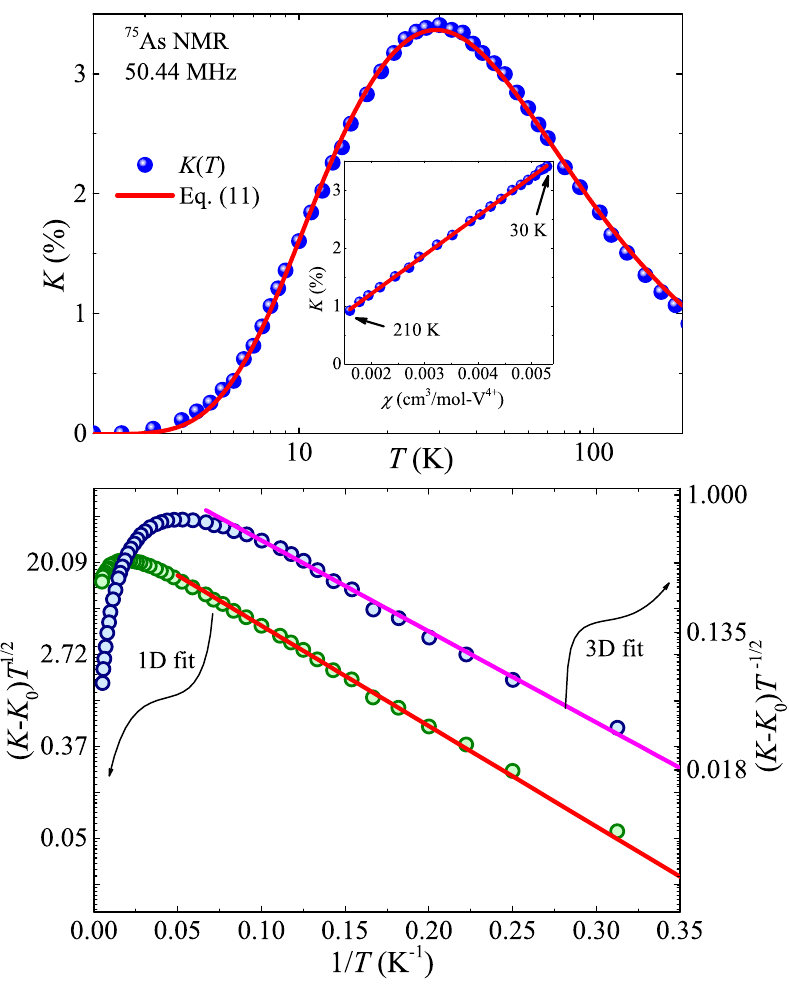}
	\caption{Upper panel: $^{75}$As NMR shift as a function of temperature. The solid line is the fit of $K(T)$ by Eq.~\eqref{K_chi}. Inset: $K$ vs $\chi$ (measured at 7~T) with temperature as an implicit parameter. The solid line is the linear fit. Lower panel: $(K-K_0)T^{1/2}$ vs $1/T$ and $(K-K_0)T^{-1/2}$ vs $1/T$ along the left and right $y$-axes, respectively. The solid lines are the fits using activation laws $(K-K_0)T^{1/2} \propto e^{-\Delta^{K}_{\rm 1D}/k_{\rm B}T}$(1D model) with $ \Delta^{K}_{\rm 1D}/k_{\rm B} \simeq 21.84 $~K and $(K-K_0)T^{-1/2} \propto e^{-\Delta^{K}_{\rm 3D}/k_{\rm B}T}$(3D model) with $ \Delta^{K}_{\rm 3D}/k_{\rm B} \simeq 13.2 $~K, respectively.}
	\label{Fig8}
\end{figure}

From Fig.~\ref{Fig7} we can clearly see that the line position is shifting with temperature. The temperature-dependent NMR shift $ K(T) $ was extracted by taking the zero quadrupole shift with respect to the reference field of GaAs and is presented in Fig.~\ref{Fig8}. Similar to $\chi(T)$, $K(T)$ also passes through a broad maximum at around 30~K, which indicates low-dimensional short-range order. At low temperatures, $K(T)$ decreases rapidly towards zero, which clearly signifies the reduction of the spin susceptibility of V$^{\rm 4+}$and the opening of a spin gap between the singlet ($S=0$) ground state and triplet ($S=1$) excited states. It also implies that the low-temperature upturn observed in $ \chi(T) $ is purely extrinsic in nature and could be due to a small amount of extrinsic impurities, defects, and/or finite crystallite size. In powder samples, the defects often break the spin chains and the unpaired spins at the end of finite chains give rise to the staggered magnetization, which also appears as low-temperature Curie tail in $ \chi(T) $.

In general, one can write $ K(T) $ in terms of $ \chi_{\rm spin} $ as
\begin{equation}\label{K_chi}
K(T) = K_0 + \frac{A_{\rm hf}}{N_{\rm A}\mu_{\rm B}}\chi_{\rm spin},
\end{equation}
where $K_0$ is the temperature-independent chemical shift and $A_{\rm hf}$ is the total hyperfine coupling between the $^{75}$As nuclei and V$^{4+}$ spins. $A_{\rm hf}$ includes contributions from transferred hyperfine coupling and the nuclear dipolar coupling, both of which are temperature-independent. The magnitude of the nuclear dipolar coupling is usually negligible compared to the transferred hyperfine coupling.
From Eq.~\eqref{K_chi}, $A_{\rm hf}$ can be calculated by taking the slope of the linear $K$ vs $\chi$ plot (inset of Fig.~\ref{Fig8}) with temperature as an implicit parameter. The $\chi(T)$ data used in the inset of Fig.~\ref{Fig8} were measured at 7~T, which is close to the field at which our NMR experiments were performed. The data for $T \geq 30$~K were fitted well to a linear function, and the slope of the fit yields $A_{\rm hf} \simeq 3.75$~T/$\mu_{\rm B}$. Nearly the same value of $A_{\rm hf}$ is reported for the isostructural compound AgVOAsO$ _4 $~\cite{Ahmed224423}. However, it is an order of magnitude larger than the one observed for $^{31}$P in 1D spin-$1/2$ chain compounds like (Sr,Ba)Cu(PO$_4$)$_2$ and K$_2$CuP$_2$O$_7$ with similar interaction pathways~\cite{Nath174436,Nath134451}. Such a large hyperfine coupling suggests a strong overlap between the $p$ orbitals of As$^{5+}$ and $d$ orbitals of V$^{4+}$ ions via the 2$p$ orbitals of O$^{2-}$. This also explains why the superexchange interaction between V$^{4+}$ ions is stronger via the V-O-As-O-V pathway than via the structural chains with the shorter V-O-V path (see Ref.~\onlinecite{Tsirlin144412} and Sec.~\ref{sec:dft} below).

In order to extract the exchange couplings, the $K(T)$ data were fitted using Eq.~\eqref{K_chi} over the whole temperature range, taking $\chi_{\rm spin}$ for the spin-$1/2$ alternating chain model given by Johnston $et.~ al.$~\cite{Johnston9558}. To minimize the number of fitting parameters, $A_{\rm hf}$ and $g$ were fixed to the values obtained from the $K$ vs $\chi$ analysis and ESR experiments, respectively. Our fit in the whole temperature range yields $K_0 \simeq 0.006 $~\%, $J/k_{\rm B} \simeq 48.14$~K, and $\alpha \simeq 0.66$. The quality of the fit was very good (see upper panel of Fig.~\ref{Fig8}) and the obtained values of $J/k_{\rm B}$ and $\alpha$ are close to those obtained from the $\chi(T)$ and $ I_{\rm ESR}(T) $ analysis. Using the above values of $J/k_{\rm B}$ and $ \alpha $, the estimated spin gap is $\sim 24.5$~K which is similar to the values reported above, yet unexpected. It should be noted that the expressions by Johnston $et.~ al.$ are applicable for the estimation of $J/k_{\rm B}$ and $\Delta/k_{\rm B}$ from the $\chi(T)$ data in the zero-field limit but our NMR experiments were carried out at a high field of 6.8~T, where the gap is expected to decrease to about $ \sim 12.3 $~K, assuming a linear decrease of the gap with field from the zero-field value $\Delta_0/k_{\rm B} \simeq 21.4$~K towards $\Delta_0/k_{\rm B}=0$ at $H_c$~\footnote{Note that from the high-field magnetization, the critical field of the gap closing is $H_{\rm c} \simeq 16$~T, and the corresponding zero-field spin gap is $\Delta/k_{\rm B} \simeq 21.4$~K. Since the $^{75}$As NMR measurements were carried out in the field of 6.8~T, the spin gap will be reduced. The reduction is estimated as $\Delta^\prime/k_{\rm B} = \frac{H g \mu_{\rm B}}{k_{\rm B}} \simeq 9.1$~K. Thus, from $^{75}$As NMR at 6.8~T one should observe a spin gap of $\Delta/k_{\rm B}-\Delta^\prime/k_{\rm B} \simeq 12.3$~K, assuming the linear field dependence of $\Delta/k_{\rm B}$.}.

One can also estimate the spin gap by analyzing the low-temperature part of the $ K(T) $ data. We fitted the low-temperature ($ T \leq 11 $~K) $ K(T) $ data by 
$K(T) = K_0+A~\chi_{\rm 1D}(T)$, where $ K_0 $ and $ A $ are arbitrary constants and $ \chi_{\rm 1D} $ is given by Eq.~\eqref{Dample} with a change that $\Delta^{\chi}_{\rm 1D}/k_{\rm B}$ is replaced by $\Delta^{K}_{\rm 1D}/k_{\rm B}$.
The obtained fitting parameters are $K_{0} \simeq 0.0052$~\%, $A \simeq 45.02$~\%, and $\Delta^{K}_{\rm 1D}/k_{\rm B} \simeq 21.8$~K. The fit is shown in the lower panel of Fig.~\ref{Fig8} where we have plotted $(K-K_{0})T^{1/2}$ vs $1/T$. The $y$-axis is shown in the log scale in order to highlight the linear behavior in the gapped region. This value of $\Delta^{K}_{\rm 1D}/k_{\rm B}$ is still higher than the expected value of $\sim 12.3$~K at $H = 6.8$~T~\footnotemark[\value{footnote}].
Note, however, that the Eq.~\eqref{Dample} used here to fit $K(T)$ is obtained in the low-field limit~\cite{Sachdev943,Damle8307} while our experimental data are taken in the field of 6.8~T, which is comparable to the thermal energy below $ \sim7 $~K. 
From the band-structure calculations, it is evident that the inter-chain couplings are non-negligible and modify the dispersion of the triplet band. Therefore, the temperature range of the fitting is expected to be dominated by the 3D correlations and, consequently, the low-temperature fit using the 1D model may not give a reliable estimation of the spin gap.

According to Ref.~\onlinecite{Taniguch2758}, for a $ d $-dimensional spin system, the low-temperature ($ k_{\rm B}T<<\Delta $) susceptibility can be expressed as
\begin{equation}\label{chi_d}
\chi_d \propto T^{(d/2)-1}\times e^{-\Delta/k_{\rm B}T}.
\end{equation}
Assuming that our low-temperature $K(T)$ data are dominated by the 3D ($ d = 3 $) correlations, the above expression is reduced to $\chi_{\rm 3D} = c T^{1/2}\times e^{-\Delta/k_{\rm B}T}$, where $ c $ is a constant. In the lower panel of Fig.~\ref{Fig8} (right $ y $-axis), we have plotted $(K - K_0) T^{-1/2}$ versus $ 1/T $, which shows a linear regime for $ T \leq 11 $~K. Our fit in this regime returns $ K_0 \simeq -0.023$~\%, $ c \simeq 1.9356$~K$^{-1/2}$, and $\Delta^{K}_{\rm 3D}/k_{\rm B}\simeq 13.2$~K. This value of $ \Delta^{K}_{\rm 3D}/k_{\rm B} $ extracted from the $ K(T) $ data using the 3D model is very close to the expected value of $\sim$12.3~K at 6.8~T~\footnotemark[\value{footnote}].

\subsubsection{Spin-lattice relaxation rate, $1/T_1$}
\begin{figure}
	\includegraphics[scale=1.09]{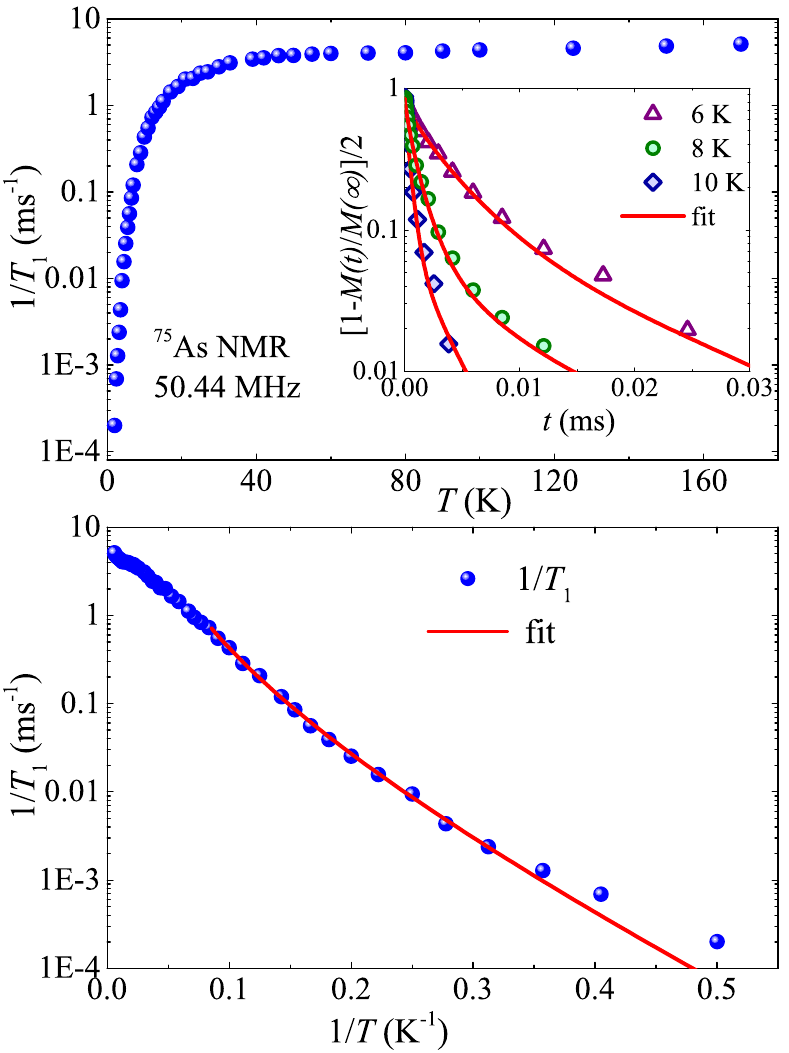}
	\caption{Upper panel: $1/T_1$ is plotted as a function of temperature. Inset: Recovery of longitudinal magnetization as a function of waiting time $t$ at three different temperatures. Solid lines are the fits using Eq.~\eqref{Double_exp}. Lower panel: $1/T_{\rm 1}$ is plotted against $1/T$. The solid line is the fit in the low-temperature region using Eq.~\eqref{T1_magnon}.}
	\label{Fig9}
\end{figure}
Spin-lattice relaxation rate, $1/T_1$, is an important parameter for understanding the dynamic properties of a spin system. It gives direct access to the low-energy spin excitations by probing the nearly zero-energy limit (in the momentum space) of the local spin-spin correlation function~\cite{Moriya23}. The $1/T_1$ measurements for the $^{75}$As nuclei were done at a field corresponding to the center of the spectra and using relatively narrow pulses. Recovery of the longitudinal magnetization at different temperatures, measured using the inversion recovery sequence, $ \pi-\tau-\pi/2 $, was fitted using the double exponential function~\cite{Gordon783,*Simmons1168}
\begin{equation}\label{Double_exp}
\frac{1}{2}\left[ 1 - \frac{M(t)}{M(\infty)}\right]  = 0.1~{\rm exp}(-t/T_1)+ 0.9~{\rm exp}(-6t/T_1),
\end{equation}
which corresponds to the $ I=3/2 $ $^{75}$As nuclei.
Here, $M(t)$ and $M(\infty)$ are the nuclear magnetizations at time $t$ and at equilibrium ($t \rightarrow \infty$), respectively. The inset of Fig.~\ref{Fig9} depicts the recovery curves at three different temperatures. The $1/T_1$ values obtained from the fit are plotted as a function of temperature in Fig.~\ref{Fig9}. At high temperatures, $ 1/T_1 $ remains constant, a typical behavior expected in the paramagnetic regime~\cite{Moriya23}. Below about 12~K, $1/T_1$ decreases rapidly towards zero because of the opening of a spin gap in the excitation spectrum.

Generally, $\frac{1}{T_1T}$ can be expressed in terms of the dynamical susceptibility $\chi(\vec{q},\omega)$ per mole of electronic spins as~\cite{Nath214430,Ranjith024422}
\begin{equation}
\frac{1}{T_{1}T} = \frac{2\gamma_{N}^{2}k_{B}}{N_{\rm A}^{2}}
\sum\limits_{\vec{q}}\mid A(\vec{q})\mid
^{2}\frac{\chi''(\vec{q},\omega_{0})}{\omega_{0}}.
\label{t1form}
\end{equation}
Here, the summation is over all the wave vectors $\vec{q}$ in the first Brillouin zone.
$A(\vec{q})$ is the form factor of the hyperfine interactions as a
function of $\vec{q}$ and
$\chi^{''}_{M}(\vec{q},\omega _{0})$ is the imaginary part of the
dynamic susceptibility at the nuclear Larmor frequency $\omega _{0}$.
When $q=0$ and $\omega_{0}=0$, the real component $\chi_{M}^{'}(\vec{q},\omega _{0})$ corresponds to the uniform static susceptibility $\chi$. In the paramagnetic regime where spins are uncorrelated, $1/T_1$ is dominated by the uniform $q=0$ fluctuations and hence $1/T_1$ remains independent of $T$~\cite{Moriya23}. In that case, $1/\chi T_1 T$ is expected to be a constant. In our compound, the temperature independent $1/T_1$ behavior at high temperatures is consistent with this expectation.

For the estimation of spin gap, we tried to fit the $1/T_1$ data by the 1D expression~\cite{Sachdev943,Damle8307}
\begin{equation}\label{T1_Dample}
1/T_{\rm 1} \propto \Delta^{T_1}_{\rm 1D} \sqrt{T} e^{-3\Delta^{T_1}_{\rm 1D}/2k_{\rm B}T}.
\end{equation}
From the fit below 11~K, the value of the spin gap is estimated to be $\Delta^{T_1}_{\rm 1D}/k_{\rm B} \simeq 18$~K, which is higher than $ \sim12.3 $~K, expected at 6.8~T~\footnotemark[\value{footnote}]. For an accurate determination of the spin gap, it would be ideal to fit the $ 1/T_1 $ data using the expression
\begin{equation}\label{T1_magnon}
1/T_{\rm 1} \propto T^{\alpha_0} \exp\left[\frac{g\mu_{\rm B}(H-H_{\rm c})}{2k_{\rm B}T}\right],
\end{equation}
which accounts for the 3D magnon excitations over the gapped region ($H < H_{\rm c}$). Here, the exponent $ \alpha_0 $ depends on the effective dimension of the magnon dispersion relation as
selected by thermal fluctuations $ k_{\rm B}T$~\cite{Mukhopadhyay177206}. On increasing $T$, $ \alpha_0 $ gradually varies from 2 (for the 3D regime, $ k_{\rm B}T  <  J_{\rm 3D} $) to 0 (for the 1D regime, $ J_{\rm 3D} \ll k_{\rm B}T  <  J $). The lower panel Fig.~\ref{Fig9} shows $1/T_1$ vs $1/T$ plot along with the fit using Eq.~\eqref{T1_magnon}, where the $y$-axis is shown in the log scale in order to highlight the activated behavior at low temperatures. Our fit in the low-$T$ region ($T\leq11$~K) with the fixed $ g = 1.99 $ (from ESR), and $H=6.8$\,T (experimental NMR field), and $ \alpha_0 =2$ (for the 3D regime) yields the critical field of the gap closing $ H_{\rm c }\simeq 17.08 $~T. The corresponding value of the zero-field spin gap is $ \Delta^{T_1}_{\rm 3D}/k_{\rm B} \simeq 22.84$~K. This value is in very good agreement with our previous estimation from the high-field magnetization data. Moreover, the value of the exponent $ \alpha_0=2 $ suggests that at low temperatures the spin-lattice is dominated by 3D correlations, as corroborated by the $\chi(T)$ analysis and band-structure calculations.

\subsection{Microscopic magnetic model}
\label{sec:dft}
Two complementary techniques can be used to estimate exchange couplings in $S=\frac12$ materials. On one hand, total-energy DFT+$U$ calculations allow a direct parametrization of the spin Hamiltonian through the so-called mapping procedure~\cite{Xiang2011}, although its results may become ambiguous and depend on computational details, such as the double-counting correction of the DFT+$U$ method~\cite{Tsirlin2011b}. On the other hand, hopping parameters $t_i$ of the uncorrelated (GGA) band structure can be directly introduced into an effective one-orbital Hubbard model, which in the half-filling regime maps onto a spin Hamiltonian for low-energy excitations. Consequently, the AFM part of the exchange is obtained as $J_i^{\rm AFM}=4t_i^2/U_{\rm eff}$, where $U_{\rm eff}\simeq 4$\,eV stands for the effective on-site Coulomb repulsion in the V $3d$ band~\cite{Tsirlin2008}. This way, relative strengths of the exchange couplings can be directly linked to the hopping parameters $t_i$, which are not biased by the choice of the Hubbard $U$ and by details of its DFT+$U$ implementation.

\begin{table}
\caption{\label{tab:exchange}
The V--V distances (in\,\r A), hopping parameters $t_i$ (in\,meV) obtained on the GGA level, and exchange couplings $J_i$ (in\,K) calculated via the DFT+$U$ mapping procedure for both NaVOAsO$_4$ and AgVOAsO$_4$. See Fig.~\ref{Fig1} for the notation of individual exchange pathways. Note that the DFT+$U$ results for AgVOAsO$_4$ differ from those in Ref.~\onlinecite{Tsirlin144412}, because the GGA functional has been used.
}
\begin{ruledtabular}
\begin{tabular}{c@{\hspace{3em}}crr@{\hspace{3em}}crr}
      & \multicolumn{3}{l}{\qquad NaVOAsO$_4$} & \multicolumn{3}{c}{AgVOAsO$_4$} \\
			& $d_{\rm V-V}$ & $t_i$ & $J_i$ & $d_{\rm V-V}$ & $t_i$ & $J_i$ \\\hline
$J$   & 5.519 & 99 &  57  & 5.587 & 87 &  45  \\
$J'$  & 5.489 & 81 &  54  & 5.556 & 71 &  41  \\
$J_a$ & 3.617 & 25 &   8  & 3.639 & 30 &   9  \\
$J_c$ & 6.073 & 9 & $-5$  & 6.118 & 3 & $-6$  \\
\end{tabular}
\end{ruledtabular}
\end{table}

Table~\ref{tab:exchange} presents the comparative DFT results for the exchange couplings in NaVOAsO$_4$ and AgVOAsO$_4$. Both compounds feature spin chains formed by $J$ and $J'$. These spin chains do not coincide with the structural chains, as shown in Fig.~\ref{Fig1}. The values of $t$ and $t'$ suggest that the intrachain couplings in NaVOAsO$_4$ are enhanced compared to the Ag analog, yet the alternation ratio given by $\alpha=(t'/t)^2\simeq 0.67$ remains nearly constant. The interchain couplings are ferromagnetic (FM) $J_c$ and AFM $J_a$, and they both become slightly weaker upon the replacement of Ag by Na. This way, NaVOAsO$_4$ is somewhat closer to the 1D alternating-chain regime than the Ag compound.

\begin{figure}
\includegraphics{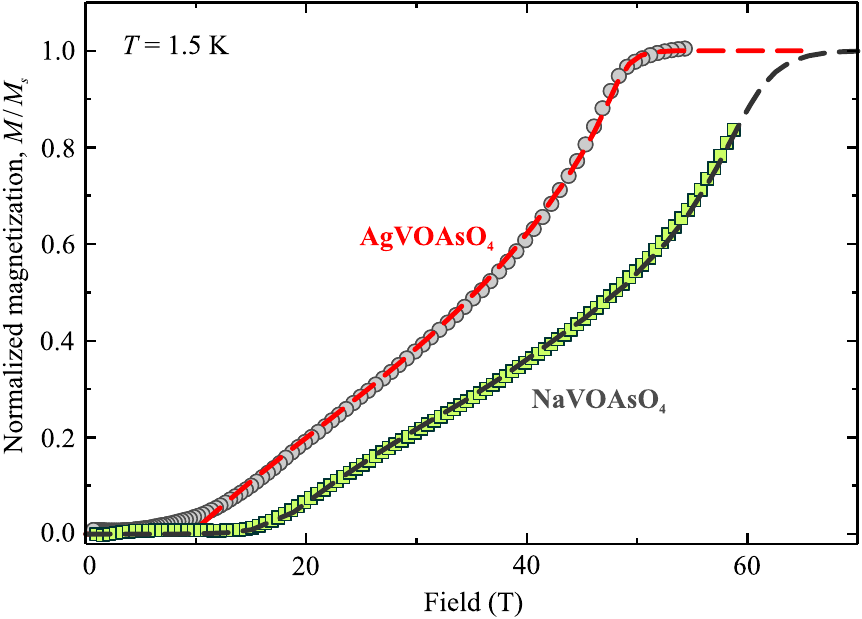}
\caption{\label{fig:mvsh}
Comparison of magnetization isotherms for AgVOAsO$_4$ and NaVOAsO$_4$ measured at 1.5\,K (paramagnetic background subtracted). Dashed lines are simulated curves for alternating spin chains uniformly coupled in 2D by an effective interchain coupling $J_{\perp}$. The fit parameters are $J=34$\,K, $\alpha=0.60$, and $J_{\perp}/J=-0.15$ for AgVOAsO$_4$ and $J=51$\,K, $\alpha=0.63$, and $J_{\perp}/J=-0.05$ for NaVOAsO$_4$.}
\end{figure}

On a more quantitative level DFT+$U$ overestimates the alternation ratio $\alpha$ and renders $J$ nearly equal to $J'$, the result not supported by the hopping parameters and also contradicting the experiment. A similar problem occurs in the case of AgVOAsO$_4$~\cite{Tsirlin144412} and may be related to the intrinsic difficulties of evaluating weak exchange couplings (or small differences between the exchange couplings) within DFT+$U$~\cite{Tsirlin2009}. All qualitative trends are, nevertheless, consistent with the experimental findings. The susceptibility fit with the alternating-chain model yields stronger exchange couplings in NaVOAsO$_4$ ($J_1\simeq 51.6$\,K, $J'\simeq 33.5$\,K) and weaker couplings in AgVOAsO$_4$ ($J_1\simeq 41.8$\,K, $J'\simeq 25.8$\,K~\cite{Tsirlin144412}) with about the same alternation ratio $\alpha\simeq 0.65$ (Na) and $\alpha\simeq 0.62$ (Ag).

For a further comparison, we take a closer look at the magnetization curves of the two compounds (Fig.~\ref{fig:mvsh}). The overall increase in the coupling energy shifts the saturation toward higher fields. AgVOAsO$_4$ saturates at 48.5\,T, whereas the saturation of NaVOAsO$_4$ is not reached even at 59\,T, the highest field of our measurement. With the 23\%\ increase in $J$ and $J'$, one expects a similar increase in $H_c$ that amounts to 10\,T in AgVOAsO$_4$ and should then be around 12\,T in NaVOAsO$_4$, much lower than the experimental value of 16\,T. This discrepancy confirms that not only the intrachain couplings increase but also the interchain couplings decrease upon the replacement of Ag by Na. Direct simulation of the magnetization curves (Fig.~\ref{fig:mvsh}) shows the best agreement for $J_{\perp}/J=-0.15$ in AgVOAsO$_4$ vs. $J_{\perp}/J=-0.05$ in NaVOAsO$_4$, thus suggesting a significant reduction in the interchain exchange~\footnote{Similar to Ref.~\onlinecite{Tsirlin144412}, we perform simulations for a 2D array of alternating spin chains uniformly coupled by an effective interchain coupling $J_{\perp}$. The actual 3D spin lattice of NaVOAsO$_4$ is frustrated and thus not amenable to QMC simulations owing to the notorious sign problem.}.

Finally, we note that individual exchange couplings in NaVOAsO$_4$ follow the general microscopic scenario for the V$^{4+}$ compounds. The distortion of the VO$_6$ octahedra puts the magnetic orbital in the plane perpendicular to the structural chains~\cite{Tsirlin144412}. This facilitates magnetic interactions between the structural chains ($J$ and $J'$) and suppresses the coupling $J_c$ along these chains. The interaction $J_a$ through a single AsO$_4$ tetrahedron is of similar strength as $J_c$ but of the opposite sign, which leads to a competing exchange scenario and the frustration of the 3D spin lattice, where the four-fold loops with one FM and three AFM couplings are present (Fig.~\ref{Fig1}). The frustrated nature of the interchain couplings was arguably the reason for inconsistencies in the fitting parameters for AgVOAsO$_4$~\cite{Tsirlin144412}: compare $J\simeq 41.8$\,K from the susceptibility fit to $J\simeq 34$\,K from the fit of the magnetization curve. This discrepancy is basically remedied in NaVOAsO$_4$, where we find a consistent fit of both susceptibility and high-field magnetization with $J\simeq 51$\,K and $\alpha=0.63-0.65$. Therefore, the frustration seems to play minor role in NaVOAsO$_4$ compared to its Ag analogue.

\section{Summary}
We have studied in detail the magnetic behavior of the quantum magnet NaVOAsO$_4$ and establish it as a new alternating spin-$1/2$ chain compound. The adjacent V$^{4+}$ ions are coupled antiferromagnetically and form alternating chains running in two crossing directions in the $ab$-plane. Compared to its Ag-analogue, the lattice parameters and unit cell volume are smaller in magnitude for NaVOAsO$_4$. As a result, the strength of the exchange interactions are found to be stronger in NaVOAsO$_4$.

The $\chi(T)$ data analysis with the alternating spin-$1/2$ chain model establishes a spin gap between the singlet ground state and the triplet excited states with the exchange couplings $J/k_{\rm B} \simeq 51.6$~K and $J^\prime/k_{\rm B} \simeq 33.6$~K, which are about 23\% larger than in the Ag-analogue.
The spin gap $\Delta_0/k_{\rm B} \simeq 21.4$~K is determined from the critical field of $ H_{\rm c} \simeq 16$~T, where the magnetization departs from zero. The relatively large $H_c$ compared to $H_c\simeq 10$\,T in AgVOAsO$_4$ is not fully accounted by the increased intrachain couplings and also reflects the reduction in the interchange couplings. Therefore, NaVOAsO$_4$ is closer to the 1D regime.

The $^{75}$As nuclei are strongly coupled to the magnetic V$^{4+}$ spins with a hyperfine coupling $A_{\rm {hf}} \simeq 3.75$~T/$\mu_{\rm B}$. The NMR shift $K(T)$ and the spin-lattice relaxation rate $1/T_{1}(T)$ show the low-temperature activated behavior, which unambiguously demonstrates the singlet ground state in this compound. The spin gap $\Delta^{\rm K}_{\rm 3D}/k_{\rm B} \simeq 13.2$~K extracted from the $ K(T) $ data using the 3D model is much closer to the expected value of $\sim 12.3$~K at 6.8~T, assuming a linear decrease of the spin gap with the applied field. Our estimated value of the zero-field spin gap $\Delta_0/k_{\rm B} \simeq 22.84$ from the $1/T_1$ analysis, accounting for the 3D magnon dispersion, is consistent with that obtained from the magnetization isotherm data, suggesting that at low temperatures NaVOAsO$_4$ acts as a 3D magnet despite its relatively weak interchain couplings. The gapped nature of the spectrum and the availability of the microscopic parameters render NaVOAsO$_4$ a model compound for high-field studies especially for exploring field-induced effects.

\acknowledgments
UA and RN would like to acknowledge BRNS, India for financial support bearing sanction No.37(3)/14/26/2017-BRNS. AAT was supported by the Federal Ministry of Education and Research through the Sofja Kovalevskaya Award of Alexander von Humboldt Foundation. We thank B.R. Sekhar for useful discussions.


\begin{thebibliography}{67}%
	\makeatletter
	\providecommand \@ifxundefined [1]{%
		\@ifx{#1\undefined}
	}%
	\providecommand \@ifnum [1]{%
		\ifnum #1\expandafter \@firstoftwo
		\else \expandafter \@secondoftwo
		\fi
	}%
	\providecommand \@ifx [1]{%
		\ifx #1\expandafter \@firstoftwo
		\else \expandafter \@secondoftwo
		\fi
	}%
	\providecommand \natexlab [1]{#1}%
	\providecommand \enquote  [1]{``#1''}%
	\providecommand \bibnamefont  [1]{#1}%
	\providecommand \bibfnamefont [1]{#1}%
	\providecommand \citenamefont [1]{#1}%
	\providecommand \href@noop [0]{\@secondoftwo}%
	\providecommand \href [0]{\begingroup \@sanitize@url \@href}%
	\providecommand \@href[1]{\@@startlink{#1}\@@href}%
	\providecommand \@@href[1]{\endgroup#1\@@endlink}%
	\providecommand \@sanitize@url [0]{\catcode `\\12\catcode `\$12\catcode
		`\&12\catcode `\#12\catcode `\^12\catcode `\_12\catcode `\%12\relax}%
	\providecommand \@@startlink[1]{}%
	\providecommand \@@endlink[0]{}%
	\providecommand \url  [0]{\begingroup\@sanitize@url \@url }%
	\providecommand \@url [1]{\endgroup\@href {#1}{\urlprefix }}%
	\providecommand \urlprefix  [0]{URL }%
	\providecommand \Eprint [0]{\href }%
	\providecommand \doibase [0]{http://dx.doi.org/}%
	\providecommand \selectlanguage [0]{\@gobble}%
	\providecommand \bibinfo  [0]{\@secondoftwo}%
	\providecommand \bibfield  [0]{\@secondoftwo}%
	\providecommand \translation [1]{[#1]}%
	\providecommand \BibitemOpen [0]{}%
	\providecommand \bibitemStop [0]{}%
	\providecommand \bibitemNoStop [0]{.\EOS\space}%
	\providecommand \EOS [0]{\spacefactor3000\relax}%
	\providecommand \BibitemShut  [1]{\csname bibitem#1\endcsname}%
	\let\auto@bib@innerbib\@empty
	\bibitem [{\citenamefont {Sachdev}(2008)}]{Sachdev173}%
	\BibitemOpen
	\bibfield  {author} {\bibinfo {author} {\bibfnamefont {S.}~\bibnamefont
			{Sachdev}},\ }\bibfield  {title} {\enquote {\bibinfo {title} {Quantum
				magnetism and criticality},}\ }\href@noop {} {\bibfield  {journal} {\bibinfo
			{journal} {Nature Phys.}\ }\textbf {\bibinfo {volume} {4}},\ \bibinfo {pages}
		{173} (\bibinfo {year} {2008})}\BibitemShut {NoStop}%
	\bibitem [{\citenamefont {Yamauchi}\ \emph {et~al.}(1999)\citenamefont
		{Yamauchi}, \citenamefont {Narumi}, \citenamefont {Kikuchi}, \citenamefont
		{Ueda}, \citenamefont {Tatani}, \citenamefont {Kobayashi}, \citenamefont
		{Kindo},\ and\ \citenamefont {Motoya}}]{Yamauchi3729}%
	\BibitemOpen
	\bibfield  {author} {\bibinfo {author} {\bibfnamefont {T.}~\bibnamefont
			{Yamauchi}}, \bibinfo {author} {\bibfnamefont {Y.}~\bibnamefont {Narumi}},
		\bibinfo {author} {\bibfnamefont {J.}~\bibnamefont {Kikuchi}}, \bibinfo
		{author} {\bibfnamefont {Y.}~\bibnamefont {Ueda}}, \bibinfo {author}
		{\bibfnamefont {K.}~\bibnamefont {Tatani}}, \bibinfo {author} {\bibfnamefont
			{T.~C.}\ \bibnamefont {Kobayashi}}, \bibinfo {author} {\bibfnamefont
			{K.}~\bibnamefont {Kindo}}, \ and\ \bibinfo {author} {\bibfnamefont
			{K.}~\bibnamefont {Motoya}},\ }\bibfield  {title} {\enquote {\bibinfo {title}
			{Two gaps in {(VO)$_2$P$_2$O$_7$}: Observation using high-field magnetization
				and {NMR}},}\ }\href {\doibase 10.1103/PhysRevLett.83.3729} {\bibfield
		{journal} {\bibinfo  {journal} {Phys. Rev. Lett.}\ }\textbf {\bibinfo
			{volume} {83}},\ \bibinfo {pages} {3729--3732} (\bibinfo {year}
		{1999})}\BibitemShut {NoStop}%
	\bibitem [{\citenamefont {Chesnut}(1966)}]{Chesnut4677}%
	\BibitemOpen
	\bibfield  {author} {\bibinfo {author} {\bibfnamefont {D.~B.}\ \bibnamefont
			{Chesnut}},\ }\bibfield  {title} {\enquote {\bibinfo {title} {Instability of
				a linear spin array: Application to {W\"urster's} blue perchlorate},}\ }\href
	{\doibase 10.1063/1.1727554} {\bibfield  {journal} {\bibinfo  {journal} {J.
				Chem. Phys.}\ }\textbf {\bibinfo {volume} {45}},\ \bibinfo {pages}
		{4677--4681} (\bibinfo {year} {1966})}\BibitemShut {NoStop}%
	\bibitem [{\citenamefont {Bray}\ \emph {et~al.}(1975)\citenamefont {Bray},
		\citenamefont {Hart}, \citenamefont {Interrante}, \citenamefont {Jacobs},
		\citenamefont {Kasper}, \citenamefont {Watkins}, \citenamefont {Wee},\ and\
		\citenamefont {Bonner}}]{Bray744}%
	\BibitemOpen
	\bibfield  {author} {\bibinfo {author} {\bibfnamefont {J.~W.}\ \bibnamefont
			{Bray}}, \bibinfo {author} {\bibfnamefont {H.~R.}\ \bibnamefont {Hart}},
		\bibinfo {author} {\bibfnamefont {L.~V.}\ \bibnamefont {Interrante}},
		\bibinfo {author} {\bibfnamefont {I.~S.}\ \bibnamefont {Jacobs}}, \bibinfo
		{author} {\bibfnamefont {J.~S.}\ \bibnamefont {Kasper}}, \bibinfo {author}
		{\bibfnamefont {G.~D.}\ \bibnamefont {Watkins}}, \bibinfo {author}
		{\bibfnamefont {S.~H.}\ \bibnamefont {Wee}}, \ and\ \bibinfo {author}
		{\bibfnamefont {J.~C.}\ \bibnamefont {Bonner}},\ }\bibfield  {title}
	{\enquote {\bibinfo {title} {Observation of a {spin-Peierls} transition in a
				{Heisenberg} antiferromagnetic linear-chain system},}\ }\href {\doibase
		10.1103/PhysRevLett.35.744} {\bibfield  {journal} {\bibinfo  {journal} {Phys.
				Rev. Lett.}\ }\textbf {\bibinfo {volume} {35}},\ \bibinfo {pages} {744--747}
		(\bibinfo {year} {1975})}\BibitemShut {NoStop}%
	\bibitem [{\citenamefont {Huizinga}\ \emph {et~al.}(1979)\citenamefont
		{Huizinga}, \citenamefont {Kommandeur}, \citenamefont {Sawatzky},
		\citenamefont {Thole}, \citenamefont {Kopinga}, \citenamefont {de~Jonge},\
		and\ \citenamefont {Roos}}]{Huizinga4723}%
	\BibitemOpen
	\bibfield  {author} {\bibinfo {author} {\bibfnamefont {S.}~\bibnamefont
			{Huizinga}}, \bibinfo {author} {\bibfnamefont {J.}~\bibnamefont
			{Kommandeur}}, \bibinfo {author} {\bibfnamefont {G.~A.}\ \bibnamefont
			{Sawatzky}}, \bibinfo {author} {\bibfnamefont {B.~T.}\ \bibnamefont {Thole}},
		\bibinfo {author} {\bibfnamefont {K.}~\bibnamefont {Kopinga}}, \bibinfo
		{author} {\bibfnamefont {W.~J.~M.}\ \bibnamefont {de~Jonge}}, \ and\ \bibinfo
		{author} {\bibfnamefont {J.}~\bibnamefont {Roos}},\ }\bibfield  {title}
	{\enquote {\bibinfo {title} {{Spin-Peierls} transition in
				{N-methyl-N-ethyl-morpholinium-ditetracyanoquinodimethanide
					[MEM-${(\mathrm{TCNQ})}_{2}$]}},}\ }\href {\doibase 10.1103/PhysRevB.19.4723}
	{\bibfield  {journal} {\bibinfo  {journal} {Phys. Rev. B}\ }\textbf {\bibinfo
			{volume} {19}},\ \bibinfo {pages} {4723--4732} (\bibinfo {year}
		{1979})}\BibitemShut {NoStop}%
	\bibitem [{\citenamefont {Hase}\ \emph {et~al.}(1993)\citenamefont {Hase},
		\citenamefont {Terasaki},\ and\ \citenamefont {Uchinokura}}]{Hase3651}%
	\BibitemOpen
	\bibfield  {author} {\bibinfo {author} {\bibfnamefont {M.}~\bibnamefont
			{Hase}}, \bibinfo {author} {\bibfnamefont {I.}~\bibnamefont {Terasaki}}, \
		and\ \bibinfo {author} {\bibfnamefont {K.}~\bibnamefont {Uchinokura}},\
	}\bibfield  {title} {\enquote {\bibinfo {title} {Observation of the
			spin-{Peierls} transition in linear {Cu$^{2+}$} (spin-1/2) chains in an
			inorganic compound {CuGeO$_3$}},}\ }\href {\doibase
	10.1103/PhysRevLett.70.3651} {\bibfield  {journal} {\bibinfo  {journal}
		{Phys. Rev. Lett.}\ }\textbf {\bibinfo {volume} {70}},\ \bibinfo {pages}
	{3651--3654} (\bibinfo {year} {1993})}\BibitemShut {NoStop}%
\bibitem [{\citenamefont {Isobe}\ and\ \citenamefont {Ueda}(1996)}]{Isobe1178}%
\BibitemOpen
\bibfield  {author} {\bibinfo {author} {\bibfnamefont {M.}~\bibnamefont
		{Isobe}}\ and\ \bibinfo {author} {\bibfnamefont {Y.}~\bibnamefont {Ueda}},\
}\bibfield  {title} {\enquote {\bibinfo {title} {Magnetic susceptibility of
		quasi-one-dimensional compound {$\alpha'$-NaV$_2$O$_5$}: Possible
		spin-{Peierls} compound with high critical temperature of {34\,K}},}\ }\href
{\doibase 10.1143/JPSJ.65.1178} {\bibfield  {journal} {\bibinfo  {journal}
		{J. Phys. Soc. Jpn.}\ }\textbf {\bibinfo {volume} {65}},\ \bibinfo {pages}
	{1178--1181} (\bibinfo {year} {1996})}\BibitemShut {NoStop}%
\bibitem [{\citenamefont {Isobe}\ \emph {et~al.}(2002)\citenamefont {Isobe},
	\citenamefont {Ninomiya}, \citenamefont {Vasil'ev},\ and\ \citenamefont
	{Ueda}}]{Isobe1423}%
\BibitemOpen
\bibfield  {author} {\bibinfo {author} {\bibfnamefont {M.}~\bibnamefont
		{Isobe}}, \bibinfo {author} {\bibfnamefont {E.}~\bibnamefont {Ninomiya}},
	\bibinfo {author} {\bibfnamefont {A.N.}\ \bibnamefont {Vasil'ev}}, \ and\
	\bibinfo {author} {\bibfnamefont {Y.}~\bibnamefont {Ueda}},\ }\bibfield
{title} {\enquote {\bibinfo {title} {Novel phase transition in spin-1/2
			linear chain systems: {NaTiSi$_2$O$_6$} and {LiTiSi$_2$O$_6$}},}\ }\href
{\doibase 10.1143/JPSJ.71.1423} {\bibfield  {journal} {\bibinfo  {journal}
		{J. Phys. Soc. Jpn.}\ }\textbf {\bibinfo {volume} {71}},\ \bibinfo {pages}
	{1423--1426} (\bibinfo {year} {2002})}\BibitemShut {NoStop}%
\bibitem [{\citenamefont {Rice}(2002)}]{Rice760}%
\BibitemOpen
\bibfield  {author} {\bibinfo {author} {\bibfnamefont {T.~M.}\ \bibnamefont
		{Rice}},\ }\bibfield  {title} {\enquote {\bibinfo {title} {To condense or not
			to condense},}\ }\href {\doibase 10.1126/science.1078819} {\bibfield
	{journal} {\bibinfo  {journal} {Science}\ }\textbf {\bibinfo {volume}
		{298}},\ \bibinfo {pages} {760--761} (\bibinfo {year} {2002})}\BibitemShut
{NoStop}%
\bibitem [{\citenamefont {Sebastian}\ \emph {et~al.}(2006)\citenamefont
	{Sebastian}, \citenamefont {Harrison}, \citenamefont {Batista}, \citenamefont
	{Balicas}, \citenamefont {Jaime}, \citenamefont {Sharma}, \citenamefont
	{Kawashima},\ and\ \citenamefont {Fisher}}]{Sebastian617}%
\BibitemOpen
\bibfield  {author} {\bibinfo {author} {\bibfnamefont {S.~E.}\ \bibnamefont
		{Sebastian}}, \bibinfo {author} {\bibfnamefont {N}~\bibnamefont {Harrison}},
	\bibinfo {author} {\bibfnamefont {C.~D.}\ \bibnamefont {Batista}}, \bibinfo
	{author} {\bibfnamefont {L.}~\bibnamefont {Balicas}}, \bibinfo {author}
	{\bibfnamefont {M.}~\bibnamefont {Jaime}}, \bibinfo {author} {\bibfnamefont
		{P.~A.}\ \bibnamefont {Sharma}}, \bibinfo {author} {\bibfnamefont
		{N}~\bibnamefont {Kawashima}}, \ and\ \bibinfo {author} {\bibfnamefont
		{I.~R.}\ \bibnamefont {Fisher}},\ }\bibfield  {title} {\enquote {\bibinfo
		{title} {Dimensional reduction at a quantum critical point},}\ }\href
{\doibase 10.1038/nature04732} {\bibfield  {journal} {\bibinfo  {journal}
		{Nature}\ }\textbf {\bibinfo {volume} {441}},\ \bibinfo {pages} {617--620}
	(\bibinfo {year} {2006})}\BibitemShut {NoStop}%
\bibitem [{\citenamefont {Aczel}\ \emph {et~al.}(2009)\citenamefont {Aczel},
	\citenamefont {Kohama}, \citenamefont {Marcenat}, \citenamefont {Weickert},
	\citenamefont {Jaime}, \citenamefont {Ayala-Valenzuela}, \citenamefont
	{McDonald}, \citenamefont {Selesnic}, \citenamefont {Dabkowska},\ and\
	\citenamefont {Luke}}]{Aczel207203}%
\BibitemOpen
\bibfield  {author} {\bibinfo {author} {\bibfnamefont {A.~A.}\ \bibnamefont
		{Aczel}}, \bibinfo {author} {\bibfnamefont {Y.}~\bibnamefont {Kohama}},
	\bibinfo {author} {\bibfnamefont {C.}~\bibnamefont {Marcenat}}, \bibinfo
	{author} {\bibfnamefont {F.}~\bibnamefont {Weickert}}, \bibinfo {author}
	{\bibfnamefont {M.}~\bibnamefont {Jaime}}, \bibinfo {author} {\bibfnamefont
		{O.~E.}\ \bibnamefont {Ayala-Valenzuela}}, \bibinfo {author} {\bibfnamefont
		{R.~D.}\ \bibnamefont {McDonald}}, \bibinfo {author} {\bibfnamefont {S.~D.}\
		\bibnamefont {Selesnic}}, \bibinfo {author} {\bibfnamefont {H.~A.}\
		\bibnamefont {Dabkowska}}, \ and\ \bibinfo {author} {\bibfnamefont {G.~M.}\
		\bibnamefont {Luke}},\ }\bibfield  {title} {\enquote {\bibinfo {title}
		{Field-induced {Bose-Einstein} condensation of triplons up to {8\,K} in
			{Sr$_3$Cr$_2$O$_8$}},}\ }\href {\doibase 10.1103/PhysRevLett.103.207203}
{\bibfield  {journal} {\bibinfo  {journal} {Phys. Rev. Lett.}\ }\textbf
	{\bibinfo {volume} {103}},\ \bibinfo {pages} {207203} (\bibinfo {year}
	{2009})}\BibitemShut {NoStop}%
\bibitem [{\citenamefont {R{\"u}egg}\ \emph {et~al.}(2003)\citenamefont
	{R{\"u}egg}, \citenamefont {Cavadini}, \citenamefont {Furrer}, \citenamefont
	{Gudel}, \citenamefont {Kramer}, \citenamefont {Mutka}, \citenamefont
	{Wildes}, \citenamefont {Habicht},\ and\ \citenamefont
	{Vorderwisch}}]{Ruegg62}%
\BibitemOpen
\bibfield  {author} {\bibinfo {author} {\bibfnamefont {Ch.}\ \bibnamefont
		{R{\"u}egg}}, \bibinfo {author} {\bibfnamefont {N}~\bibnamefont {Cavadini}},
	\bibinfo {author} {\bibfnamefont {A}~\bibnamefont {Furrer}}, \bibinfo
	{author} {\bibfnamefont {H.-U.}\ \bibnamefont {Gudel}}, \bibinfo {author}
	{\bibfnamefont {K.}~\bibnamefont {Kramer}}, \bibinfo {author} {\bibfnamefont
		{H.}~\bibnamefont {Mutka}}, \bibinfo {author} {\bibfnamefont
		{A.}~\bibnamefont {Wildes}}, \bibinfo {author} {\bibfnamefont
		{K.}~\bibnamefont {Habicht}}, \ and\ \bibinfo {author} {\bibfnamefont
		{P.}~\bibnamefont {Vorderwisch}},\ }\bibfield  {title} {\enquote {\bibinfo
		{title} {{Bose-Einstein} condensation of the triplet states in the magnetic
			insulator {TlCuCl$_3$}},}\ }\href {\doibase 10.1038/nature01617} {\bibfield
	{journal} {\bibinfo  {journal} {Nature}\ }\textbf {\bibinfo {volume} {423}},\
	\bibinfo {pages} {62--65} (\bibinfo {year} {2003})}\BibitemShut {NoStop}%
\bibitem [{\citenamefont {Zapf}\ \emph {et~al.}(2014)\citenamefont {Zapf},
	\citenamefont {Jaime},\ and\ \citenamefont {Batista}}]{Zapf2014}%
\BibitemOpen
\bibfield  {author} {\bibinfo {author} {\bibfnamefont {V.}~\bibnamefont
		{Zapf}}, \bibinfo {author} {\bibfnamefont {M.}~\bibnamefont {Jaime}}, \ and\
	\bibinfo {author} {\bibfnamefont {C.}~\bibnamefont {Batista}},\ }\bibfield
{title} {\enquote {\bibinfo {title} {{Bose-Einstein} condensation in quantum
			magnets},}\ }\href {\doibase 10.1103/RevModPhys.86.563} {\bibfield  {journal}
	{\bibinfo  {journal} {Rev. Mod. Phys.}\ }\textbf {\bibinfo {volume} {86}},\
	\bibinfo {pages} {563--614} (\bibinfo {year} {2014})}\BibitemShut {NoStop}%
\bibitem [{\citenamefont {Freitas}\ \emph {et~al.}(2017)\citenamefont
	{Freitas}, \citenamefont {Alves},\ and\ \citenamefont
	{Paduan-Filho}}]{Freitas184426}%
\BibitemOpen
\bibfield  {author} {\bibinfo {author} {\bibfnamefont {R.~S.}\ \bibnamefont
		{Freitas}}, \bibinfo {author} {\bibfnamefont {W.~A.}\ \bibnamefont {Alves}},
	\ and\ \bibinfo {author} {\bibfnamefont {A.}~\bibnamefont {Paduan-Filho}},\
}\bibfield  {title} {\enquote {\bibinfo {title} {Magnetic-field-induced
		ordered phase in the chloro-bridged copper(ii) dimer system
		$[\mathrm{C}{\mathrm{u}}_{2}{(\mathrm{apyhist})}_{2}\mathrm{C}{\mathrm{l}}_{2}]{(\mathrm{Cl}{\mathrm{O}}_{4})}_{2}$},}\
}\href {\doibase 10.1103/PhysRevB.95.184426} {\bibfield  {journal} {\bibinfo
	{journal} {Phys. Rev. B}\ }\textbf {\bibinfo {volume} {95}},\ \bibinfo
{pages} {184426} (\bibinfo {year} {2017})}\BibitemShut {NoStop}%
\bibitem [{\citenamefont {Tomonaga}(1950)}]{Tomonaga544}%
\BibitemOpen
\bibfield  {author} {\bibinfo {author} {\bibfnamefont {S.}~\bibnamefont
		{Tomonaga}},\ }\bibfield  {title} {\enquote {\bibinfo {title} {Remarks on
			{Bloch's} method of sound waves applied to many-fermion problems},}\ }\href
{\doibase 10.1143/ptp/5.4.544} {\bibfield  {journal} {\bibinfo  {journal}
		{Prog. Theor. Phys.}\ }\textbf {\bibinfo {volume} {5}},\ \bibinfo {pages}
	{544--569} (\bibinfo {year} {1950})}\BibitemShut {NoStop}%
\bibitem [{\citenamefont {Luttinger}(2014)}]{Luttinger3}%
\BibitemOpen
\bibfield  {author} {\bibinfo {author} {\bibfnamefont {J.~M.}\ \bibnamefont
		{Luttinger}},\ }\href@noop {} {\emph {\bibinfo {title} {Luttinger Model: The
			First 50 Years and Some New Directions}}}\ (\bibinfo  {publisher} {World
	Scientific},\ \bibinfo {year} {2014})\ pp.\ \bibinfo {pages}
{3--11}\BibitemShut {NoStop}%
\bibitem [{\citenamefont {Jeong}\ \emph {et~al.}(2017)\citenamefont {Jeong},
	\citenamefont {Mayaffre}, \citenamefont {Berthier}, \citenamefont
	{Schmidiger}, \citenamefont {Zheludev},\ and\ \citenamefont
	{Horvati\ifmmode~\acute{c}\else \'{c}\fi{}}}]{Jeong167206}%
\BibitemOpen
\bibfield  {author} {\bibinfo {author} {\bibfnamefont {M.}~\bibnamefont
		{Jeong}}, \bibinfo {author} {\bibfnamefont {H.}~\bibnamefont {Mayaffre}},
	\bibinfo {author} {\bibfnamefont {C.}~\bibnamefont {Berthier}}, \bibinfo
	{author} {\bibfnamefont {D.}~\bibnamefont {Schmidiger}}, \bibinfo {author}
	{\bibfnamefont {A.}~\bibnamefont {Zheludev}}, \ and\ \bibinfo {author}
	{\bibfnamefont {M.}~\bibnamefont {Horvati\ifmmode~\acute{c}\else
			\'{c}\fi{}}},\ }\bibfield  {title} {\enquote {\bibinfo {title}
		{Magnetic-order crossover in coupled spin ladders},}\ }\href {\doibase
	10.1103/PhysRevLett.118.167206} {\bibfield  {journal} {\bibinfo  {journal}
		{Phys. Rev. Lett.}\ }\textbf {\bibinfo {volume} {118}},\ \bibinfo {pages}
	{167206} (\bibinfo {year} {2017})}\BibitemShut {NoStop}%
\bibitem [{\citenamefont {Hong}\ \emph {et~al.}(2010)\citenamefont {Hong},
	\citenamefont {Kim}, \citenamefont {Hotta}, \citenamefont {Takano},
	\citenamefont {Tremelling}, \citenamefont {Turnbull}, \citenamefont {Landee},
	\citenamefont {Kang}, \citenamefont {Christensen}, \citenamefont {Lefmann},
	\citenamefont {Schmidt}, \citenamefont {Uhrig},\ and\ \citenamefont
	{Broholm}}]{Hong137207}%
\BibitemOpen
\bibfield  {author} {\bibinfo {author} {\bibfnamefont {T.}~\bibnamefont
		{Hong}}, \bibinfo {author} {\bibfnamefont {Y.~H.}\ \bibnamefont {Kim}},
	\bibinfo {author} {\bibfnamefont {C.}~\bibnamefont {Hotta}}, \bibinfo
	{author} {\bibfnamefont {Y.}~\bibnamefont {Takano}}, \bibinfo {author}
	{\bibfnamefont {G.}~\bibnamefont {Tremelling}}, \bibinfo {author}
	{\bibfnamefont {M.~M.}\ \bibnamefont {Turnbull}}, \bibinfo {author}
	{\bibfnamefont {C.~P.}\ \bibnamefont {Landee}}, \bibinfo {author}
	{\bibfnamefont {H.-J.}\ \bibnamefont {Kang}}, \bibinfo {author}
	{\bibfnamefont {N.~B.}\ \bibnamefont {Christensen}}, \bibinfo {author}
	{\bibfnamefont {K.}~\bibnamefont {Lefmann}}, \bibinfo {author} {\bibfnamefont
		{K.~P.}\ \bibnamefont {Schmidt}}, \bibinfo {author} {\bibfnamefont {G.~S.}\
		\bibnamefont {Uhrig}}, \ and\ \bibinfo {author} {\bibfnamefont
		{C.}~\bibnamefont {Broholm}},\ }\bibfield  {title} {\enquote {\bibinfo
		{title} {Field-induced {Tomonaga-Luttinger} liquid phase of a two-leg
			spin-1/2 ladder with strong leg interactions},}\ }\href {\doibase
	10.1103/PhysRevLett.105.137207} {\bibfield  {journal} {\bibinfo  {journal}
		{Phys. Rev. Lett.}\ }\textbf {\bibinfo {volume} {105}},\ \bibinfo {pages}
	{137207} (\bibinfo {year} {2010})}\BibitemShut {NoStop}%
\bibitem [{\citenamefont {Povarov}\ \emph {et~al.}(2015)\citenamefont
	{Povarov}, \citenamefont {Schmidiger}, \citenamefont {Reynolds},
	\citenamefont {Bewley},\ and\ \citenamefont {Zheludev}}]{Povarov2015}%
\BibitemOpen
\bibfield  {author} {\bibinfo {author} {\bibfnamefont {K.~Yu.}\ \bibnamefont
		{Povarov}}, \bibinfo {author} {\bibfnamefont {D.}~\bibnamefont {Schmidiger}},
	\bibinfo {author} {\bibfnamefont {N.}~\bibnamefont {Reynolds}}, \bibinfo
	{author} {\bibfnamefont {R.}~\bibnamefont {Bewley}}, \ and\ \bibinfo {author}
	{\bibfnamefont {A.}~\bibnamefont {Zheludev}},\ }\bibfield  {title} {\enquote
	{\bibinfo {title} {Scaling of temporal correlations in an attractive
			{Tomonaga-Luttinger} spin liquid},}\ }\href {\doibase
	10.1103/PhysRevB.91.020406} {\bibfield  {journal} {\bibinfo  {journal} {Phys.
			Rev. B}\ }\textbf {\bibinfo {volume} {91}},\ \bibinfo {pages} {020406(R)}
	(\bibinfo {year} {2015})}\BibitemShut {NoStop}%
\bibitem [{\citenamefont {Jeong}\ \emph {et~al.}(2016)\citenamefont {Jeong},
	\citenamefont {Schmidiger}, \citenamefont {Mayaffre}, \citenamefont {Klanj{\u
			s}ek}, \citenamefont {Berthier}, \citenamefont {Knafo}, \citenamefont
	{Ballon}, \citenamefont {Vignolle}, \citenamefont {Kr\"amer}, \citenamefont
	{Zheludev},\ and\ \citenamefont {Horvati{\'c}}}]{Jeong2016}%
\BibitemOpen
\bibfield  {author} {\bibinfo {author} {\bibfnamefont {M.}~\bibnamefont
		{Jeong}}, \bibinfo {author} {\bibfnamefont {D.}~\bibnamefont {Schmidiger}},
	\bibinfo {author} {\bibfnamefont {H.}~\bibnamefont {Mayaffre}}, \bibinfo
	{author} {\bibfnamefont {M.}~\bibnamefont {Klanj{\u s}ek}}, \bibinfo {author}
	{\bibfnamefont {C.}~\bibnamefont {Berthier}}, \bibinfo {author}
	{\bibfnamefont {W.}~\bibnamefont {Knafo}}, \bibinfo {author} {\bibfnamefont
		{G.}~\bibnamefont {Ballon}}, \bibinfo {author} {\bibfnamefont
		{B.}~\bibnamefont {Vignolle}}, \bibinfo {author} {\bibfnamefont
		{S.}~\bibnamefont {Kr\"amer}}, \bibinfo {author} {\bibfnamefont
		{A.}~\bibnamefont {Zheludev}}, \ and\ \bibinfo {author} {\bibfnamefont
		{M.}~\bibnamefont {Horvati{\'c}}},\ }\bibfield  {title} {\enquote {\bibinfo
		{title} {Dichotomy between attractive and repulsive {Tomonaga-Luttinger}
			liquids in spin ladders},}\ }\href {\doibase 10.1103/PhysRevLett.117.106402}
{\bibfield  {journal} {\bibinfo  {journal} {Phys. Rev. Lett.}\ }\textbf
	{\bibinfo {volume} {117}},\ \bibinfo {pages} {106402} (\bibinfo {year}
	{2016})}\BibitemShut {NoStop}%
\bibitem [{\citenamefont {M\"oller}\ \emph {et~al.}(2017)\citenamefont
	{M\"oller}, \citenamefont {Lancaster}, \citenamefont {Blundell},
	\citenamefont {Pratt}, \citenamefont {Baker}, \citenamefont {Xiao},
	\citenamefont {Williams}, \citenamefont {Hayes}, \citenamefont {Turnbull},\
	and\ \citenamefont {Landee}}]{Moeller2017}%
\BibitemOpen
\bibfield  {author} {\bibinfo {author} {\bibfnamefont {J.~S.}\ \bibnamefont
		{M\"oller}}, \bibinfo {author} {\bibfnamefont {T.}~\bibnamefont {Lancaster}},
	\bibinfo {author} {\bibfnamefont {S.~J.}\ \bibnamefont {Blundell}}, \bibinfo
	{author} {\bibfnamefont {F.~L.}\ \bibnamefont {Pratt}}, \bibinfo {author}
	{\bibfnamefont {P.~J.}\ \bibnamefont {Baker}}, \bibinfo {author}
	{\bibfnamefont {F.}~\bibnamefont {Xiao}}, \bibinfo {author} {\bibfnamefont
		{R.~C.}\ \bibnamefont {Williams}}, \bibinfo {author} {\bibfnamefont
		{W.}~\bibnamefont {Hayes}}, \bibinfo {author} {\bibfnamefont {M.~M.}\
		\bibnamefont {Turnbull}}, \ and\ \bibinfo {author} {\bibfnamefont {C.~P.}\
		\bibnamefont {Landee}},\ }\bibfield  {title} {\enquote {\bibinfo {title}
		{Quantum-critical spin dynamics in a {Tomonaga-Luttinger} liquid studied with
			muon-spin relaxation},}\ }\href {\doibase 10.1103/PhysRevB.95.020402}
{\bibfield  {journal} {\bibinfo  {journal} {Phys. Rev. B}\ }\textbf {\bibinfo
		{volume} {95}},\ \bibinfo {pages} {020402(R)} (\bibinfo {year}
	{2017})}\BibitemShut {NoStop}%
\bibitem [{\citenamefont {Thielemann}\ \emph {et~al.}(2009)\citenamefont
	{Thielemann}, \citenamefont {R\"uegg}, \citenamefont {Kiefer}, \citenamefont
	{R\o{}nnow}, \citenamefont {Normand}, \citenamefont {Bouillot}, \citenamefont
	{Kollath}, \citenamefont {Orignac}, \citenamefont {Citro}, \citenamefont
	{Giamarchi}, \citenamefont {L\"auchli}, \citenamefont {Biner}, \citenamefont
	{Kr\"amer}, \citenamefont {Wolff-Fabris}, \citenamefont {Zapf}, \citenamefont
	{Jaime}, \citenamefont {Stahn}, \citenamefont {Christensen}, \citenamefont
	{Grenier}, \citenamefont {McMorrow},\ and\ \citenamefont
	{Mesot}}]{Thielemann020408}%
\BibitemOpen
\bibfield  {author} {\bibinfo {author} {\bibfnamefont {B.}~\bibnamefont
		{Thielemann}}, \bibinfo {author} {\bibfnamefont {Ch.}\ \bibnamefont
		{R\"uegg}}, \bibinfo {author} {\bibfnamefont {K.}~\bibnamefont {Kiefer}},
	\bibinfo {author} {\bibfnamefont {H.~M.}\ \bibnamefont {R\o{}nnow}}, \bibinfo
	{author} {\bibfnamefont {B.}~\bibnamefont {Normand}}, \bibinfo {author}
	{\bibfnamefont {P.}~\bibnamefont {Bouillot}}, \bibinfo {author}
	{\bibfnamefont {C.}~\bibnamefont {Kollath}}, \bibinfo {author} {\bibfnamefont
		{E.}~\bibnamefont {Orignac}}, \bibinfo {author} {\bibfnamefont
		{R.}~\bibnamefont {Citro}}, \bibinfo {author} {\bibfnamefont
		{T.}~\bibnamefont {Giamarchi}}, \bibinfo {author} {\bibfnamefont {A.~M.}\
		\bibnamefont {L\"auchli}}, \bibinfo {author} {\bibfnamefont {D.}~\bibnamefont
		{Biner}}, \bibinfo {author} {\bibfnamefont {K.~W.}\ \bibnamefont {Kr\"amer}},
	\bibinfo {author} {\bibfnamefont {F.}~\bibnamefont {Wolff-Fabris}}, \bibinfo
	{author} {\bibfnamefont {V.~S.}\ \bibnamefont {Zapf}}, \bibinfo {author}
	{\bibfnamefont {M.}~\bibnamefont {Jaime}}, \bibinfo {author} {\bibfnamefont
		{J.}~\bibnamefont {Stahn}}, \bibinfo {author} {\bibfnamefont {N.~B.}\
		\bibnamefont {Christensen}}, \bibinfo {author} {\bibfnamefont
		{B.}~\bibnamefont {Grenier}}, \bibinfo {author} {\bibfnamefont {D.~F.}\
		\bibnamefont {McMorrow}}, \ and\ \bibinfo {author} {\bibfnamefont
		{J.}~\bibnamefont {Mesot}},\ }\bibfield  {title} {\enquote {\bibinfo {title}
		{Field-controlled magnetic order in the quantum spin-ladder system
			{(Hpip)$_2$CuBr$_4$}},}\ }\href {\doibase 10.1103/PhysRevB.79.020408}
{\bibfield  {journal} {\bibinfo  {journal} {Phys. Rev. B}\ }\textbf {\bibinfo
		{volume} {79}},\ \bibinfo {pages} {020408} (\bibinfo {year}
	{2009})}\BibitemShut {NoStop}%
\bibitem [{\citenamefont {R\"uegg}\ \emph {et~al.}(2008)\citenamefont
	{R\"uegg}, \citenamefont {Kiefer}, \citenamefont {Thielemann}, \citenamefont
	{McMorrow}, \citenamefont {Zapf}, \citenamefont {Normand}, \citenamefont
	{Zvonarev}, \citenamefont {Bouillot}, \citenamefont {Kollath}, \citenamefont
	{Giamarchi}, \citenamefont {Capponi}, \citenamefont {Poilblanc},
	\citenamefont {Biner},\ and\ \citenamefont {Kr\"amer}}]{Ruegg247202}%
\BibitemOpen
\bibfield  {author} {\bibinfo {author} {\bibfnamefont {Ch.}\ \bibnamefont
		{R\"uegg}}, \bibinfo {author} {\bibfnamefont {K.}~\bibnamefont {Kiefer}},
	\bibinfo {author} {\bibfnamefont {B.}~\bibnamefont {Thielemann}}, \bibinfo
	{author} {\bibfnamefont {D.~F.}\ \bibnamefont {McMorrow}}, \bibinfo {author}
	{\bibfnamefont {V.}~\bibnamefont {Zapf}}, \bibinfo {author} {\bibfnamefont
		{B.}~\bibnamefont {Normand}}, \bibinfo {author} {\bibfnamefont {M.~B.}\
		\bibnamefont {Zvonarev}}, \bibinfo {author} {\bibfnamefont {P.}~\bibnamefont
		{Bouillot}}, \bibinfo {author} {\bibfnamefont {C.}~\bibnamefont {Kollath}},
	\bibinfo {author} {\bibfnamefont {T.}~\bibnamefont {Giamarchi}}, \bibinfo
	{author} {\bibfnamefont {S.}~\bibnamefont {Capponi}}, \bibinfo {author}
	{\bibfnamefont {D.}~\bibnamefont {Poilblanc}}, \bibinfo {author}
	{\bibfnamefont {D.}~\bibnamefont {Biner}}, \ and\ \bibinfo {author}
	{\bibfnamefont {K.~W.}\ \bibnamefont {Kr\"amer}},\ }\bibfield  {title}
{\enquote {\bibinfo {title} {Thermodynamics of the spin {Luttinger} liquid in
			a model ladder material},}\ }\href {\doibase 10.1103/PhysRevLett.101.247202}
{\bibfield  {journal} {\bibinfo  {journal} {Phys. Rev. Lett.}\ }\textbf
	{\bibinfo {volume} {101}},\ \bibinfo {pages} {247202} (\bibinfo {year}
	{2008})}\BibitemShut {NoStop}%
\bibitem [{\citenamefont {Orignac}\ \emph {et~al.}(2007)\citenamefont
	{Orignac}, \citenamefont {Citro},\ and\ \citenamefont
	{Giamarchi}}]{Orignac140403}%
\BibitemOpen
\bibfield  {author} {\bibinfo {author} {\bibfnamefont {E.}~\bibnamefont
		{Orignac}}, \bibinfo {author} {\bibfnamefont {R.}~\bibnamefont {Citro}}, \
	and\ \bibinfo {author} {\bibfnamefont {T.}~\bibnamefont {Giamarchi}},\
}\bibfield  {title} {\enquote {\bibinfo {title} {Critical properties and
		{Bose-Einstein} condensation in dimer spin systems},}\ }\href {\doibase
10.1103/PhysRevB.75.140403} {\bibfield  {journal} {\bibinfo  {journal} {Phys.
		Rev. B}\ }\textbf {\bibinfo {volume} {75}},\ \bibinfo {pages} {140403}
(\bibinfo {year} {2007})}\BibitemShut {NoStop}%
\bibitem [{\citenamefont {Willenberg}\ \emph {et~al.}(2015)\citenamefont
	{Willenberg}, \citenamefont {Ryll}, \citenamefont {Kiefer}, \citenamefont
	{Tennant}, \citenamefont {Groitl}, \citenamefont {Rolfs}, \citenamefont
	{Manuel}, \citenamefont {Khalyavin}, \citenamefont {Rule}, \citenamefont
	{Wolter},\ and\ \citenamefont {S\"ullow}}]{Willenberg060407}%
\BibitemOpen
\bibfield  {author} {\bibinfo {author} {\bibfnamefont {B.}~\bibnamefont
		{Willenberg}}, \bibinfo {author} {\bibfnamefont {H.}~\bibnamefont {Ryll}},
	\bibinfo {author} {\bibfnamefont {K.}~\bibnamefont {Kiefer}}, \bibinfo
	{author} {\bibfnamefont {D.~A.}\ \bibnamefont {Tennant}}, \bibinfo {author}
	{\bibfnamefont {F.}~\bibnamefont {Groitl}}, \bibinfo {author} {\bibfnamefont
		{K.}~\bibnamefont {Rolfs}}, \bibinfo {author} {\bibfnamefont
		{P.}~\bibnamefont {Manuel}}, \bibinfo {author} {\bibfnamefont
		{D.}~\bibnamefont {Khalyavin}}, \bibinfo {author} {\bibfnamefont {K.~C.}\
		\bibnamefont {Rule}}, \bibinfo {author} {\bibfnamefont {A.~U.~B.}\
		\bibnamefont {Wolter}}, \ and\ \bibinfo {author} {\bibfnamefont
		{S.}~\bibnamefont {S\"ullow}},\ }\bibfield  {title} {\enquote {\bibinfo
		{title} {Luttinger liquid behavior in the alternating spin-chain system
			copper nitrate},}\ }\href {\doibase 10.1103/PhysRevB.91.060407} {\bibfield
	{journal} {\bibinfo  {journal} {Phys. Rev. B}\ }\textbf {\bibinfo {volume}
		{91}},\ \bibinfo {pages} {060407} (\bibinfo {year} {2015})}\BibitemShut
{NoStop}%
\bibitem [{\citenamefont {Kim}\ and\ \citenamefont
	{Aronson}(2013)}]{Kim017201}%
\BibitemOpen
\bibfield  {author} {\bibinfo {author} {\bibfnamefont {M.~S.}\ \bibnamefont
		{Kim}}\ and\ \bibinfo {author} {\bibfnamefont {M.~C.}\ \bibnamefont
		{Aronson}},\ }\bibfield  {title} {\enquote {\bibinfo {title} {Spin liquids
			and antiferromagnetic order in the shastry-sutherland-lattice compound
			${\mathrm{yb}}_{2}{\mathrm{pt}}_{2}\mathrm{Pb}$},}\ }\href {\doibase
	10.1103/PhysRevLett.110.017201} {\bibfield  {journal} {\bibinfo  {journal}
		{Phys. Rev. Lett.}\ }\textbf {\bibinfo {volume} {110}},\ \bibinfo {pages}
	{017201} (\bibinfo {year} {2013})}\BibitemShut {NoStop}%
\bibitem [{\citenamefont {Kodama}\ \emph {et~al.}(2002)\citenamefont {Kodama},
	\citenamefont {Takigawa}, \citenamefont {Horvati{\'c}}, \citenamefont
	{Berthier}, \citenamefont {Kageyama}, \citenamefont {Ueda}, \citenamefont
	{Miyahara}, \citenamefont {Becca},\ and\ \citenamefont {Mila}}]{Kodama395}%
\BibitemOpen
\bibfield  {author} {\bibinfo {author} {\bibfnamefont {K.}~\bibnamefont
		{Kodama}}, \bibinfo {author} {\bibfnamefont {M.}~\bibnamefont {Takigawa}},
	\bibinfo {author} {\bibfnamefont {M.}~\bibnamefont {Horvati{\'c}}}, \bibinfo
	{author} {\bibfnamefont {C.}~\bibnamefont {Berthier}}, \bibinfo {author}
	{\bibfnamefont {H.}~\bibnamefont {Kageyama}}, \bibinfo {author}
	{\bibfnamefont {Y.}~\bibnamefont {Ueda}}, \bibinfo {author} {\bibfnamefont
		{S.}~\bibnamefont {Miyahara}}, \bibinfo {author} {\bibfnamefont
		{F.}~\bibnamefont {Becca}}, \ and\ \bibinfo {author} {\bibfnamefont
		{F.}~\bibnamefont {Mila}},\ }\bibfield  {title} {\enquote {\bibinfo {title}
		{Magnetic superstructure in the two-dimensional quantum antiferromagnet
			{SrCu$_2$(BO$_3)_2$}},}\ }\href {\doibase 10.1126/science.1075045} {\bibfield
	{journal} {\bibinfo  {journal} {Science}\ }\textbf {\bibinfo {volume}
		{298}},\ \bibinfo {pages} {395--399} (\bibinfo {year} {2002})}\BibitemShut
{NoStop}%
\bibitem [{\citenamefont {Tsirlin}\ \emph
	{et~al.}(2011{\natexlab{a}})\citenamefont {Tsirlin}, \citenamefont {Nath},
	\citenamefont {Sichelschmidt}, \citenamefont {Skourski}, \citenamefont
	{Geibel},\ and\ \citenamefont {Rosner}}]{Tsirlin144412}%
\BibitemOpen
\bibfield  {author} {\bibinfo {author} {\bibfnamefont {A.A.}\ \bibnamefont
		{Tsirlin}}, \bibinfo {author} {\bibfnamefont {R.}~\bibnamefont {Nath}},
	\bibinfo {author} {\bibfnamefont {J.}~\bibnamefont {Sichelschmidt}}, \bibinfo
	{author} {\bibfnamefont {Y.}~\bibnamefont {Skourski}}, \bibinfo {author}
	{\bibfnamefont {C.}~\bibnamefont {Geibel}}, \ and\ \bibinfo {author}
	{\bibfnamefont {H.}~\bibnamefont {Rosner}},\ }\bibfield  {title} {\enquote
	{\bibinfo {title} {Frustrated couplings between alternating
			spin-$\frac{1}{2}$ chains in {AgVOAsO$_4$}},}\ }\href {\doibase
	10.1103/PhysRevB.83.144412} {\bibfield  {journal} {\bibinfo  {journal} {Phys.
			Rev. B}\ }\textbf {\bibinfo {volume} {83}},\ \bibinfo {pages} {144412}
	(\bibinfo {year} {2011}{\natexlab{a}})}\BibitemShut {NoStop}%
\bibitem [{\citenamefont {Ahmed}\ \emph {et~al.}(2017)\citenamefont {Ahmed},
	\citenamefont {Khuntia}, \citenamefont {Ranjith}, \citenamefont {Rosner},
	\citenamefont {Baenitz}, \citenamefont {Tsirlin},\ and\ \citenamefont
	{Nath}}]{Ahmed224423}%
\BibitemOpen
\bibfield  {author} {\bibinfo {author} {\bibfnamefont {N.}~\bibnamefont
		{Ahmed}}, \bibinfo {author} {\bibfnamefont {P.}~\bibnamefont {Khuntia}},
	\bibinfo {author} {\bibfnamefont {K.~M.}\ \bibnamefont {Ranjith}}, \bibinfo
	{author} {\bibfnamefont {H.}~\bibnamefont {Rosner}}, \bibinfo {author}
	{\bibfnamefont {M.}~\bibnamefont {Baenitz}}, \bibinfo {author} {\bibfnamefont
		{A.~A.}\ \bibnamefont {Tsirlin}}, \ and\ \bibinfo {author} {\bibfnamefont
		{R.}~\bibnamefont {Nath}},\ }\bibfield  {title} {\enquote {\bibinfo {title}
		{Alternating spin chain compound {AgVOAsO$_4$} probed by {$^{75}$As NMR}},}\
}\href {\doibase 10.1103/PhysRevB.96.224423} {\bibfield  {journal} {\bibinfo
	{journal} {Phys. Rev. B}\ }\textbf {\bibinfo {volume} {96}},\ \bibinfo
{pages} {224423} (\bibinfo {year} {2017})}\BibitemShut {NoStop}%
\bibitem [{\citenamefont {Haddad}\ \emph {et~al.}(1992)\citenamefont {Haddad},
	\citenamefont {Jouini},\ and\ \citenamefont {Piffard}}]{Haddad57}%
\BibitemOpen
\bibfield  {author} {\bibinfo {author} {\bibfnamefont {A.}~\bibnamefont
		{Haddad}}, \bibinfo {author} {\bibfnamefont {T.}~\bibnamefont {Jouini}}, \
	and\ \bibinfo {author} {\bibfnamefont {Y.}~\bibnamefont {Piffard}},\
}\bibfield  {title} {\enquote {\bibinfo {title} {Preparation and crystal
		structure of {NaVOAsO$_4$}},}\ }\href@noop {} {\bibfield  {journal} {\bibinfo
	{journal} {Eur. J. Solid State Inorg. Chem.}\ }\textbf {\bibinfo {volume}
	{29}},\ \bibinfo {pages} {57--63} (\bibinfo {year} {1992})}\BibitemShut
{NoStop}%
\bibitem [{\citenamefont {Rodr{\'\i}guez-Carvajal}(1993)}]{Carvajal55}%
\BibitemOpen
\bibfield  {author} {\bibinfo {author} {\bibfnamefont {J.}~\bibnamefont
		{Rodr{\'\i}guez-Carvajal}},\ }\bibfield  {title} {\enquote {\bibinfo {title}
		{Recent advances in magnetic structure determination by neutron powder
			diffraction},}\ }\href {\doibase
	http://dx.doi.org/10.1016/0921-4526(93)90108-I} {\bibfield  {journal}
	{\bibinfo  {journal} {Physica B: Condens. Matter}\ }\textbf {\bibinfo
		{volume} {192}},\ \bibinfo {pages} {55 -- 69} (\bibinfo {year}
	{1993})}\BibitemShut {NoStop}%
\bibitem [{\citenamefont {Tsirlin}\ \emph {et~al.}(2009)\citenamefont
	{Tsirlin}, \citenamefont {Schmidt}, \citenamefont {Skourski}, \citenamefont
	{Nath}, \citenamefont {Geibel},\ and\ \citenamefont
	{Rosner}}]{Tsirlin132407}%
\BibitemOpen
\bibfield  {author} {\bibinfo {author} {\bibfnamefont {A.A.}\ \bibnamefont
		{Tsirlin}}, \bibinfo {author} {\bibfnamefont {B.}~\bibnamefont {Schmidt}},
	\bibinfo {author} {\bibfnamefont {Y.}~\bibnamefont {Skourski}}, \bibinfo
	{author} {\bibfnamefont {R.}~\bibnamefont {Nath}}, \bibinfo {author}
	{\bibfnamefont {C.}~\bibnamefont {Geibel}}, \ and\ \bibinfo {author}
	{\bibfnamefont {H.}~\bibnamefont {Rosner}},\ }\bibfield  {title} {\enquote
	{\bibinfo {title} {Exploring the spin-$\frac{1}{2}$ frustrated square lattice
			model with high-field magnetization studies},}\ }\href {\doibase
	10.1103/PhysRevB.80.132407} {\bibfield  {journal} {\bibinfo  {journal} {Phys.
			Rev. B}\ }\textbf {\bibinfo {volume} {80}},\ \bibinfo {pages} {132407}
	(\bibinfo {year} {2009})}\BibitemShut {NoStop}%
\bibitem [{\citenamefont {Koepernik}\ and\ \citenamefont
	{Eschrig}(1999)}]{fplo}%
\BibitemOpen
\bibfield  {author} {\bibinfo {author} {\bibfnamefont {K.}~\bibnamefont
		{Koepernik}}\ and\ \bibinfo {author} {\bibfnamefont {H.}~\bibnamefont
		{Eschrig}},\ }\bibfield  {title} {\enquote {\bibinfo {title} {Full-potential
			nonorthogonal local-orbital minimum-basis band-structure scheme},}\ }\href
{\doibase 10.1103/PhysRevB.59.1743} {\bibfield  {journal} {\bibinfo
		{journal} {Phys. Rev. B}\ }\textbf {\bibinfo {volume} {59}},\ \bibinfo
	{pages} {1743--1757} (\bibinfo {year} {1999})}\BibitemShut {NoStop}%
\bibitem [{\citenamefont {Perdew}\ \emph {et~al.}(1996)\citenamefont {Perdew},
	\citenamefont {Burke},\ and\ \citenamefont {Ernzerhof}}]{pbe96}%
\BibitemOpen
\bibfield  {author} {\bibinfo {author} {\bibfnamefont {J.~P.}\ \bibnamefont
		{Perdew}}, \bibinfo {author} {\bibfnamefont {K.}~\bibnamefont {Burke}}, \
	and\ \bibinfo {author} {\bibfnamefont {M.}~\bibnamefont {Ernzerhof}},\
}\bibfield  {title} {\enquote {\bibinfo {title} {Generalized gradient
		approximation made simple},}\ }\href {\doibase 10.1103/PhysRevLett.77.3865}
{\bibfield  {journal} {\bibinfo  {journal} {Phys. Rev. Lett.}\ }\textbf
	{\bibinfo {volume} {77}},\ \bibinfo {pages} {3865--3868} (\bibinfo {year}
	{1996})}\BibitemShut {NoStop}%
\bibitem [{\citenamefont {Nath}\ \emph
	{et~al.}(2008{\natexlab{a}})\citenamefont {Nath}, \citenamefont {Tsirlin},
	\citenamefont {Kaul}, \citenamefont {Baenitz}, \citenamefont {B\"uttgen},
	\citenamefont {Geibel},\ and\ \citenamefont {Rosner}}]{Nath024418}%
\BibitemOpen
\bibfield  {author} {\bibinfo {author} {\bibfnamefont {R.}~\bibnamefont
		{Nath}}, \bibinfo {author} {\bibfnamefont {A.~A.}\ \bibnamefont {Tsirlin}},
	\bibinfo {author} {\bibfnamefont {E.~E.}\ \bibnamefont {Kaul}}, \bibinfo
	{author} {\bibfnamefont {M.}~\bibnamefont {Baenitz}}, \bibinfo {author}
	{\bibfnamefont {N.}~\bibnamefont {B\"uttgen}}, \bibinfo {author}
	{\bibfnamefont {C.}~\bibnamefont {Geibel}}, \ and\ \bibinfo {author}
	{\bibfnamefont {H.}~\bibnamefont {Rosner}},\ }\bibfield  {title} {\enquote
	{\bibinfo {title} {Strong frustration due to competing ferromagnetic and
			antiferromagnetic interactions: Magnetic properties of {M(VO)$_2$(PO$_4)_2$}
			{(M = Ca and Sr)}},}\ }\href {\doibase 10.1103/PhysRevB.78.024418} {\bibfield
	{journal} {\bibinfo  {journal} {Phys. Rev. B}\ }\textbf {\bibinfo {volume}
		{78}},\ \bibinfo {pages} {024418} (\bibinfo {year}
	{2008}{\natexlab{a}})}\BibitemShut {NoStop}%
\bibitem [{\citenamefont {Tsirlin}\ and\ \citenamefont
	{Rosner}(2011)}]{Tsirlin2011c}%
\BibitemOpen
\bibfield  {author} {\bibinfo {author} {\bibfnamefont {A.~A.}\ \bibnamefont
		{Tsirlin}}\ and\ \bibinfo {author} {\bibfnamefont {H.}~\bibnamefont
		{Rosner}},\ }\bibfield  {title} {\enquote {\bibinfo {title} {Ab initio
			modeling of {Bose-Einstein} condensation in {Pb$_2$V$_3$O$_9$}},}\ }\href
{\doibase 10.1103/PhysRevB.83.064415} {\bibfield  {journal} {\bibinfo
		{journal} {Phys. Rev. B}\ }\textbf {\bibinfo {volume} {83}},\ \bibinfo
	{pages} {064415} (\bibinfo {year} {2011})}\BibitemShut {NoStop}%
\bibitem [{\citenamefont {Todo}\ and\ \citenamefont {Kato}(2001)}]{loop}%
\BibitemOpen
\bibfield  {author} {\bibinfo {author} {\bibfnamefont {S.}~\bibnamefont
		{Todo}}\ and\ \bibinfo {author} {\bibfnamefont {K.}~\bibnamefont {Kato}},\
}\bibfield  {title} {\enquote {\bibinfo {title} {Cluster algorithms for
		{general-$S$} quantum spin systems},}\ }\href@noop {} {\bibfield  {journal}
{\bibinfo  {journal} {Phys. Rev. Lett.}\ }\textbf {\bibinfo {volume} {87}},\
\bibinfo {pages} {047203} (\bibinfo {year} {2001})}\BibitemShut {NoStop}%
\bibitem [{\citenamefont {Alet}\ \emph {et~al.}(2005)\citenamefont {Alet},
	\citenamefont {Wessel},\ and\ \citenamefont {Troyer}}]{dirloop}%
\BibitemOpen
\bibfield  {author} {\bibinfo {author} {\bibfnamefont {F.}~\bibnamefont
		{Alet}}, \bibinfo {author} {\bibfnamefont {S.}~\bibnamefont {Wessel}}, \ and\
	\bibinfo {author} {\bibfnamefont {M.}~\bibnamefont {Troyer}},\ }\bibfield
{title} {\enquote {\bibinfo {title} {Generalized directed loop method for
			quantum {Monte Carlo} simulations},}\ }\href@noop {} {\bibfield  {journal}
	{\bibinfo  {journal} {Phys. Rev. E}\ }\textbf {\bibinfo {volume} {71}},\
	\bibinfo {pages} {036706} (\bibinfo {year} {2005})},\ \bibinfo {note} {and
	references therein}\BibitemShut {NoStop}%
\bibitem [{\citenamefont {Albuquerque}\ \emph {et~al.}(2007)\citenamefont
	{Albuquerque}, \citenamefont {Alet}, \citenamefont {Corboz}, \citenamefont
	{Dayal}, \citenamefont {Feiguin}, \citenamefont {Fuchs}, \citenamefont
	{Gamper}, \citenamefont {Gull}, \citenamefont {G\"urtler}, \citenamefont
	{Honecker}, \citenamefont {Igarashi}, \citenamefont {K\"orner}, \citenamefont
	{Kozhevnikov}, \citenamefont {L\"auchli}, \citenamefont {Manmana},
	\citenamefont {Matsumoto}, \citenamefont {McCulloch}, \citenamefont {Michel},
	\citenamefont {Noack}, \citenamefont {Paw{\l}owski}, \citenamefont {Pollet},
	\citenamefont {Pruschke}, \citenamefont {Schollw\"ock}, \citenamefont {Todo},
	\citenamefont {Trebst}, \citenamefont {Troyer}, \citenamefont {Werner},\ and\
	\citenamefont {Wessel}}]{alps}%
\BibitemOpen
\bibfield  {author} {\bibinfo {author} {\bibfnamefont {A.F.}\ \bibnamefont
		{Albuquerque}}, \bibinfo {author} {\bibfnamefont {F.}~\bibnamefont {Alet}},
	\bibinfo {author} {\bibfnamefont {P.}~\bibnamefont {Corboz}}, \bibinfo
	{author} {\bibfnamefont {P.}~\bibnamefont {Dayal}}, \bibinfo {author}
	{\bibfnamefont {A.}~\bibnamefont {Feiguin}}, \bibinfo {author} {\bibfnamefont
		{S.}~\bibnamefont {Fuchs}}, \bibinfo {author} {\bibfnamefont
		{L.}~\bibnamefont {Gamper}}, \bibinfo {author} {\bibfnamefont
		{E.}~\bibnamefont {Gull}}, \bibinfo {author} {\bibfnamefont {S.}~\bibnamefont
		{G\"urtler}}, \bibinfo {author} {\bibfnamefont {A.}~\bibnamefont {Honecker}},
	\bibinfo {author} {\bibfnamefont {R.}~\bibnamefont {Igarashi}}, \bibinfo
	{author} {\bibfnamefont {M.}~\bibnamefont {K\"orner}}, \bibinfo {author}
	{\bibfnamefont {A.}~\bibnamefont {Kozhevnikov}}, \bibinfo {author}
	{\bibfnamefont {A.}~\bibnamefont {L\"auchli}}, \bibinfo {author}
	{\bibfnamefont {S.R.}\ \bibnamefont {Manmana}}, \bibinfo {author}
	{\bibfnamefont {M.}~\bibnamefont {Matsumoto}}, \bibinfo {author}
	{\bibfnamefont {I.P.}\ \bibnamefont {McCulloch}}, \bibinfo {author}
	{\bibfnamefont {F.}~\bibnamefont {Michel}}, \bibinfo {author} {\bibfnamefont
		{R.M.}\ \bibnamefont {Noack}}, \bibinfo {author} {\bibfnamefont
		{G.}~\bibnamefont {Paw{\l}owski}}, \bibinfo {author} {\bibfnamefont
		{L.}~\bibnamefont {Pollet}}, \bibinfo {author} {\bibfnamefont
		{T.}~\bibnamefont {Pruschke}}, \bibinfo {author} {\bibfnamefont
		{U.}~\bibnamefont {Schollw\"ock}}, \bibinfo {author} {\bibfnamefont
		{S.}~\bibnamefont {Todo}}, \bibinfo {author} {\bibfnamefont {S.}~\bibnamefont
		{Trebst}}, \bibinfo {author} {\bibfnamefont {M.}~\bibnamefont {Troyer}},
	\bibinfo {author} {\bibfnamefont {P.}~\bibnamefont {Werner}}, \ and\ \bibinfo
	{author} {\bibfnamefont {S.}~\bibnamefont {Wessel}},\ }\bibfield  {title}
{\enquote {\bibinfo {title} {The {ALPS} project release 1.3: Open-source
			software for strongly correlated systems},}\ }\href@noop {} {\bibfield
	{journal} {\bibinfo  {journal} {J. Magn. Magn. Mater.}\ }\textbf {\bibinfo
		{volume} {310}},\ \bibinfo {pages} {1187--1193} (\bibinfo {year}
	{2007})}\BibitemShut {NoStop}%
\bibitem [{\citenamefont {Hirota}\ \emph {et~al.}(1994)\citenamefont {Hirota},
	\citenamefont {Cox}, \citenamefont {Lorenzo}, \citenamefont {Shirane},
	\citenamefont {Tranquada}, \citenamefont {Hase}, \citenamefont {Uchinokura},
	\citenamefont {Kojima}, \citenamefont {Shibuya},\ and\ \citenamefont
	{Tanaka}}]{Hirota736}%
\BibitemOpen
\bibfield  {author} {\bibinfo {author} {\bibfnamefont {K.}~\bibnamefont
		{Hirota}}, \bibinfo {author} {\bibfnamefont {D.~E.}\ \bibnamefont {Cox}},
	\bibinfo {author} {\bibfnamefont {J.~E.}\ \bibnamefont {Lorenzo}}, \bibinfo
	{author} {\bibfnamefont {G.}~\bibnamefont {Shirane}}, \bibinfo {author}
	{\bibfnamefont {J.~M.}\ \bibnamefont {Tranquada}}, \bibinfo {author}
	{\bibfnamefont {M.}~\bibnamefont {Hase}}, \bibinfo {author} {\bibfnamefont
		{K.}~\bibnamefont {Uchinokura}}, \bibinfo {author} {\bibfnamefont
		{H.}~\bibnamefont {Kojima}}, \bibinfo {author} {\bibfnamefont
		{Y.}~\bibnamefont {Shibuya}}, \ and\ \bibinfo {author} {\bibfnamefont
		{I.}~\bibnamefont {Tanaka}},\ }\bibfield  {title} {\enquote {\bibinfo {title}
		{Dimerization of {CuGeO$_3$} in the spin-{Peierls} state},}\ }\href {\doibase
	10.1103/PhysRevLett.73.736} {\bibfield  {journal} {\bibinfo  {journal} {Phys.
			Rev. Lett.}\ }\textbf {\bibinfo {volume} {73}},\ \bibinfo {pages} {736--739}
	(\bibinfo {year} {1994})}\BibitemShut {NoStop}%
\bibitem [{\citenamefont {Fujii}\ \emph {et~al.}(1997)\citenamefont {Fujii},
	\citenamefont {Nakao}, \citenamefont {Yosihama}, \citenamefont {Nishi},
	\citenamefont {Nakajima}, \citenamefont {Kakurai}, \citenamefont {Isobe},
	\citenamefont {Ueda},\ and\ \citenamefont {Sawa}}]{Fuji326}%
\BibitemOpen
\bibfield  {author} {\bibinfo {author} {\bibfnamefont {Y.}~\bibnamefont
		{Fujii}}, \bibinfo {author} {\bibfnamefont {H.}~\bibnamefont {Nakao}},
	\bibinfo {author} {\bibfnamefont {T.}~\bibnamefont {Yosihama}}, \bibinfo
	{author} {\bibfnamefont {M.}~\bibnamefont {Nishi}}, \bibinfo {author}
	{\bibfnamefont {K.}~\bibnamefont {Nakajima}}, \bibinfo {author}
	{\bibfnamefont {K.}~\bibnamefont {Kakurai}}, \bibinfo {author} {\bibfnamefont
		{M.}~\bibnamefont {Isobe}}, \bibinfo {author} {\bibfnamefont
		{Y.}~\bibnamefont {Ueda}}, \ and\ \bibinfo {author} {\bibfnamefont
		{H.}~\bibnamefont {Sawa}},\ }\bibfield  {title} {\enquote {\bibinfo {title}
		{New inorganic spin-{Peierls} compound {NaV$_2$O$_5$} evidenced by x-ray and
			neutron scattering},}\ }\href {\doibase 10.1143/JPSJ.66.326} {\bibfield
	{journal} {\bibinfo  {journal} {J. Phys. Soc. Jpn.}\ }\textbf {\bibinfo
		{volume} {66}},\ \bibinfo {pages} {326--329} (\bibinfo {year}
	{1997})}\BibitemShut {NoStop}%
\bibitem [{\citenamefont {Wallace}(1998)}]{Wallace1998}%
\BibitemOpen
\bibfield  {author} {\bibinfo {author} {\bibfnamefont {D.C.}\ \bibnamefont
		{Wallace}},\ }\href@noop {} {\emph {\bibinfo {title} {Thermodynamics of
			crystals}}}\ (\bibinfo  {publisher} {Dover},\ \bibinfo {address} {New York},\
\bibinfo {year} {1998})\BibitemShut {NoStop}%
\bibitem [{\citenamefont {Bag}\ \emph {et~al.}(2018)\citenamefont {Bag},
	\citenamefont {Baral},\ and\ \citenamefont {Nath}}]{Pallab2018}%
\BibitemOpen
\bibfield  {author} {\bibinfo {author} {\bibfnamefont {P.}~\bibnamefont
		{Bag}}, \bibinfo {author} {\bibfnamefont {P.~R.}\ \bibnamefont {Baral}}, \
	and\ \bibinfo {author} {\bibfnamefont {R.}~\bibnamefont {Nath}},\ }\bibfield
{title} {\enquote {\bibinfo {title} {Cluster spin-glass behavior and memory
			effect in {Cr$_{0.5}$Fe$_{0.5}$Ga}},}\ }\href@noop {} {\bibfield  {journal}
	{\bibinfo  {journal} {Phys. Rev. B (Submitted)}\ } (\bibinfo {year}
	{2018})}\BibitemShut {NoStop}%
\bibitem [{\citenamefont {Arjun}\ \emph {et~al.}(2017)\citenamefont {Arjun},
	\citenamefont {Kumar}, \citenamefont {Anjana}, \citenamefont {Thirumurugan},
	\citenamefont {Sichelschmidt}, \citenamefont {Mahajan},\ and\ \citenamefont
	{Nath}}]{Arjun174421}%
\BibitemOpen
\bibfield  {author} {\bibinfo {author} {\bibfnamefont {U.}~\bibnamefont
		{Arjun}}, \bibinfo {author} {\bibfnamefont {V.}~\bibnamefont {Kumar}},
	\bibinfo {author} {\bibfnamefont {P.~K.}\ \bibnamefont {Anjana}}, \bibinfo
	{author} {\bibfnamefont {A.}~\bibnamefont {Thirumurugan}}, \bibinfo {author}
	{\bibfnamefont {J.}~\bibnamefont {Sichelschmidt}}, \bibinfo {author}
	{\bibfnamefont {A.~V.}\ \bibnamefont {Mahajan}}, \ and\ \bibinfo {author}
	{\bibfnamefont {R.}~\bibnamefont {Nath}},\ }\bibfield  {title} {\enquote
	{\bibinfo {title} {Singlet ground state in the spin-$\frac{1}{2}$ weakly
			coupled dimer compound
			{NH$_4$[(V$_2$O$_3)_2$(4,4'-bpy)$_2$(H$_2$PO$_4$)(PO$_4)_2]\!\cdot\!0.5$H$_2$O}},}\
}\href {\doibase 10.1103/PhysRevB.95.174421} {\bibfield  {journal} {\bibinfo
	{journal} {Phys. Rev. B}\ }\textbf {\bibinfo {volume} {95}},\ \bibinfo
{pages} {174421} (\bibinfo {year} {2017})}\BibitemShut {NoStop}%
\bibitem [{\citenamefont {Johnston}\ \emph {et~al.}(2000)\citenamefont
	{Johnston}, \citenamefont {Kremer}, \citenamefont {Troyer}, \citenamefont
	{Wang}, \citenamefont {Kl\"umper}, \citenamefont {Bud'ko}, \citenamefont
	{Panchula},\ and\ \citenamefont {Canfield}}]{Johnston9558}%
\BibitemOpen
\bibfield  {author} {\bibinfo {author} {\bibfnamefont {D.~C.}\ \bibnamefont
		{Johnston}}, \bibinfo {author} {\bibfnamefont {R.~K.}\ \bibnamefont
		{Kremer}}, \bibinfo {author} {\bibfnamefont {M.}~\bibnamefont {Troyer}},
	\bibinfo {author} {\bibfnamefont {X.}~\bibnamefont {Wang}}, \bibinfo {author}
	{\bibfnamefont {A.}~\bibnamefont {Kl\"umper}}, \bibinfo {author}
	{\bibfnamefont {S.~L.}\ \bibnamefont {Bud'ko}}, \bibinfo {author}
	{\bibfnamefont {A.~F.}\ \bibnamefont {Panchula}}, \ and\ \bibinfo {author}
	{\bibfnamefont {P.~C.}\ \bibnamefont {Canfield}},\ }\bibfield  {title}
{\enquote {\bibinfo {title} {Thermodynamics of spin $s=1/2$ antiferromagnetic
			uniform and alternating-exchange {Heisenberg} chains},}\ }\href {\doibase
	10.1103/PhysRevB.61.9558} {\bibfield  {journal} {\bibinfo  {journal} {Phys.
			Rev. B}\ }\textbf {\bibinfo {volume} {61}},\ \bibinfo {pages} {9558--9606}
	(\bibinfo {year} {2000})}\BibitemShut {NoStop}%
\bibitem [{\citenamefont {Sachdev}\ and\ \citenamefont
	{Damle}(1997)}]{Sachdev943}%
\BibitemOpen
\bibfield  {author} {\bibinfo {author} {\bibfnamefont {S.}~\bibnamefont
		{Sachdev}}\ and\ \bibinfo {author} {\bibfnamefont {K.}~\bibnamefont
		{Damle}},\ }\bibfield  {title} {\enquote {\bibinfo {title} {Low temperature
			spin diffusion in the one-dimensional quantum {$O(3)$} nonlinear
			$\mathit{\ensuremath{\sigma}}$ model},}\ }\href {\doibase
	10.1103/PhysRevLett.78.943} {\bibfield  {journal} {\bibinfo  {journal} {Phys.
			Rev. Lett.}\ }\textbf {\bibinfo {volume} {78}},\ \bibinfo {pages} {943--946}
	(\bibinfo {year} {1997})}\BibitemShut {NoStop}%
\bibitem [{\citenamefont {Yogi}\ \emph {et~al.}(2015)\citenamefont {Yogi},
	\citenamefont {Ahmed}, \citenamefont {Nath}, \citenamefont {Tsirlin},
	\citenamefont {Kundu}, \citenamefont {Mahajan}, \citenamefont
	{Sichelschmidt}, \citenamefont {Roy},\ and\ \citenamefont
	{Furukawa}}]{Yogi024413}%
\BibitemOpen
\bibfield  {author} {\bibinfo {author} {\bibfnamefont {A.}~\bibnamefont
		{Yogi}}, \bibinfo {author} {\bibfnamefont {N.}~\bibnamefont {Ahmed}},
	\bibinfo {author} {\bibfnamefont {R.}~\bibnamefont {Nath}}, \bibinfo {author}
	{\bibfnamefont {A.~A.}\ \bibnamefont {Tsirlin}}, \bibinfo {author}
	{\bibfnamefont {S.}~\bibnamefont {Kundu}}, \bibinfo {author} {\bibfnamefont
		{A.~V.}\ \bibnamefont {Mahajan}}, \bibinfo {author} {\bibfnamefont
		{J.}~\bibnamefont {Sichelschmidt}}, \bibinfo {author} {\bibfnamefont
		{B.}~\bibnamefont {Roy}}, \ and\ \bibinfo {author} {\bibfnamefont
		{Y.}~\bibnamefont {Furukawa}},\ }\bibfield  {title} {\enquote {\bibinfo
		{title} {Antiferromagnetism of {Zn$_2$VO(PO$_4)_2$} and the dilution with
			{Ti$^{4+}$}},}\ }\href {\doibase 10.1103/PhysRevB.91.024413} {\bibfield
	{journal} {\bibinfo  {journal} {Phys. Rev. B}\ }\textbf {\bibinfo {volume}
		{91}},\ \bibinfo {pages} {024413} (\bibinfo {year} {2015})}\BibitemShut
{NoStop}%
\bibitem [{\citenamefont {Curro}(2009)}]{Curro026502}%
\BibitemOpen
\bibfield  {author} {\bibinfo {author} {\bibfnamefont {N.J.}\ \bibnamefont
		{Curro}},\ }\bibfield  {title} {\enquote {\bibinfo {title} {Nuclear magnetic
			resonance in the heavy fermion superconductors},}\ }\href
{http://stacks.iop.org/0034-4885/72/i=2/a=026502} {\bibfield  {journal}
	{\bibinfo  {journal} {Rep. Prog. Phys.}\ }\textbf {\bibinfo {volume} {72}},\
	\bibinfo {pages} {026502} (\bibinfo {year} {2009})}\BibitemShut {NoStop}%
\bibitem [{\citenamefont {Slichter}(1992)}]{Slichter1992}%
\BibitemOpen
\bibfield  {author} {\bibinfo {author} {\bibfnamefont {C.~P.}\ \bibnamefont
		{Slichter}},\ }\href@noop {} {\emph {\bibinfo {title} {Principle of Nuclear
			Magnetic Resonance}}},\ \bibinfo {edition} {3rd}\ ed.\ (\bibinfo  {publisher}
{Springer},\ \bibinfo {address} {New York},\ \bibinfo {year}
{1992})\BibitemShut {NoStop}%
\bibitem [{\citenamefont {Grafe}\ \emph {et~al.}(2008)\citenamefont {Grafe},
	\citenamefont {Paar}, \citenamefont {Lang}, \citenamefont {Curro},
	\citenamefont {Behr}, \citenamefont {Werner}, \citenamefont {Hamann-Borrero},
	\citenamefont {Hess}, \citenamefont {Leps}, \citenamefont {Klingeler},\ and\
	\citenamefont {B\"uchner}}]{Grafe047003}%
\BibitemOpen
\bibfield  {author} {\bibinfo {author} {\bibfnamefont {H.-J.}\ \bibnamefont
		{Grafe}}, \bibinfo {author} {\bibfnamefont {D.}~\bibnamefont {Paar}},
	\bibinfo {author} {\bibfnamefont {G.}~\bibnamefont {Lang}}, \bibinfo {author}
	{\bibfnamefont {N.~J.}\ \bibnamefont {Curro}}, \bibinfo {author}
	{\bibfnamefont {G.}~\bibnamefont {Behr}}, \bibinfo {author} {\bibfnamefont
		{J.}~\bibnamefont {Werner}}, \bibinfo {author} {\bibfnamefont
		{J.}~\bibnamefont {Hamann-Borrero}}, \bibinfo {author} {\bibfnamefont
		{C.}~\bibnamefont {Hess}}, \bibinfo {author} {\bibfnamefont {N.}~\bibnamefont
		{Leps}}, \bibinfo {author} {\bibfnamefont {R.}~\bibnamefont {Klingeler}}, \
	and\ \bibinfo {author} {\bibfnamefont {B.}~\bibnamefont {B\"uchner}},\
}\bibfield  {title} {\enquote {\bibinfo {title} {{$^{75}\mathrm{As}$ NMR}
		studies of superconducting {LaFeAsO$_{0.9}$F$_{0.1}$}},}\ }\href {\doibase
10.1103/PhysRevLett.101.047003} {\bibfield  {journal} {\bibinfo  {journal}
	{Phys. Rev. Lett.}\ }\textbf {\bibinfo {volume} {101}},\ \bibinfo {pages}
{047003} (\bibinfo {year} {2008})}\BibitemShut {NoStop}%
\bibitem [{\citenamefont {Fagot-Revurat}\ \emph {et~al.}(2000)\citenamefont
	{Fagot-Revurat}, \citenamefont {Mehring},\ and\ \citenamefont
	{Kremer}}]{Revurat4176}%
\BibitemOpen
\bibfield  {author} {\bibinfo {author} {\bibfnamefont {Y.}~\bibnamefont
		{Fagot-Revurat}}, \bibinfo {author} {\bibfnamefont {M.}~\bibnamefont
		{Mehring}}, \ and\ \bibinfo {author} {\bibfnamefont {R.~K.}\ \bibnamefont
		{Kremer}},\ }\bibfield  {title} {\enquote {\bibinfo {title}
		{Charge-order-driven spin-{Peierls} transition in
			{$\alpha'$-NaV$_2$O$_5$}},}\ }\href {\doibase 10.1103/PhysRevLett.84.4176}
{\bibfield  {journal} {\bibinfo  {journal} {Phys. Rev. Lett.}\ }\textbf
	{\bibinfo {volume} {84}},\ \bibinfo {pages} {4176--4179} (\bibinfo {year}
	{2000})}\BibitemShut {NoStop}%
\bibitem [{\citenamefont {Nath}\ \emph {et~al.}(2005)\citenamefont {Nath},
	\citenamefont {Mahajan}, \citenamefont {B\"uttgen}, \citenamefont {Kegler},
	\citenamefont {Loidl},\ and\ \citenamefont {Bobroff}}]{Nath174436}%
\BibitemOpen
\bibfield  {author} {\bibinfo {author} {\bibfnamefont {R.}~\bibnamefont
		{Nath}}, \bibinfo {author} {\bibfnamefont {A.~V.}\ \bibnamefont {Mahajan}},
	\bibinfo {author} {\bibfnamefont {N.}~\bibnamefont {B\"uttgen}}, \bibinfo
	{author} {\bibfnamefont {C.}~\bibnamefont {Kegler}}, \bibinfo {author}
	{\bibfnamefont {A.}~\bibnamefont {Loidl}}, \ and\ \bibinfo {author}
	{\bibfnamefont {J.}~\bibnamefont {Bobroff}},\ }\bibfield  {title} {\enquote
	{\bibinfo {title} {Study of one-dimensional nature of {$S=1∕2$}
			{(Sr,Ba)$_2$Cu(PO$_4)_2$} and {BaCuP$_2$O$_7$} via {$^{31}\mathrm{P}$
				NMR}},}\ }\href {\doibase 10.1103/PhysRevB.71.174436} {\bibfield  {journal}
	{\bibinfo  {journal} {Phys. Rev. B}\ }\textbf {\bibinfo {volume} {71}},\
	\bibinfo {pages} {174436} (\bibinfo {year} {2005})}\BibitemShut {NoStop}%
\bibitem [{\citenamefont {Nath}\ \emph
	{et~al.}(2008{\natexlab{b}})\citenamefont {Nath}, \citenamefont {Kasinathan},
	\citenamefont {Rosner}, \citenamefont {Baenitz},\ and\ \citenamefont
	{Geibel}}]{Nath134451}%
\BibitemOpen
\bibfield  {author} {\bibinfo {author} {\bibfnamefont {R.}~\bibnamefont
		{Nath}}, \bibinfo {author} {\bibfnamefont {D.}~\bibnamefont {Kasinathan}},
	\bibinfo {author} {\bibfnamefont {H.}~\bibnamefont {Rosner}}, \bibinfo
	{author} {\bibfnamefont {M.}~\bibnamefont {Baenitz}}, \ and\ \bibinfo
	{author} {\bibfnamefont {C.}~\bibnamefont {Geibel}},\ }\bibfield  {title}
{\enquote {\bibinfo {title} {Electronic and magnetic properties of
			{K$_2$CuP$_2$O$_7$}: A model {$S=\frac{1}{2}$ Heisenberg} chain system},}\
}\href {\doibase 10.1103/PhysRevB.77.134451} {\bibfield  {journal} {\bibinfo
	{journal} {Phys. Rev. B}\ }\textbf {\bibinfo {volume} {77}},\ \bibinfo
{pages} {134451} (\bibinfo {year} {2008}{\natexlab{b}})}\BibitemShut
{NoStop}%
\bibitem [{Note1()}]{Note1}%
\BibitemOpen
\bibinfo {note} {Note that from the high-field magnetization, the critical
	field of the gap closing is $H_{\protect \rm c} \simeq 16$~T, and the
	corresponding zero-field spin gap is $\Delta /k_{\protect \rm B} \simeq
	21.4$~K. Since the $^{75}$As NMR measurements were carried out in the field
	of 6.8~T, the spin gap will be reduced. The reduction is estimated as $\Delta
	^\prime /k_{\protect \rm B} = \protect \frac {H g \mu _{\protect \rm
			B}}{k_{\protect \rm B}} \simeq 9.1$~K. Thus, from $^{75}$As NMR at 6.8~T one
	should observe a spin gap of $\Delta /k_{\protect \rm B}-\Delta ^\prime
	/k_{\protect \rm B} \simeq 12.3$~K, assuming the linear field dependence of
	$\Delta /k_{\protect \rm B}$.}\BibitemShut {Stop}%
\bibitem [{\citenamefont {Damle}\ and\ \citenamefont
	{Sachdev}(1998)}]{Damle8307}%
\BibitemOpen
\bibfield  {author} {\bibinfo {author} {\bibfnamefont {K.}~\bibnamefont
		{Damle}}\ and\ \bibinfo {author} {\bibfnamefont {S.}~\bibnamefont
		{Sachdev}},\ }\bibfield  {title} {\enquote {\bibinfo {title} {Spin dynamics
			and transport in gapped one-dimensional {Heisenberg} antiferromagnets at
			nonzero temperatures},}\ }\href {\doibase 10.1103/PhysRevB.57.8307}
{\bibfield  {journal} {\bibinfo  {journal} {Phys. Rev. B}\ }\textbf {\bibinfo
		{volume} {57}},\ \bibinfo {pages} {8307--8339} (\bibinfo {year}
	{1998})}\BibitemShut {NoStop}%
\bibitem [{\citenamefont {Taniguchi}\ \emph {et~al.}(1995)\citenamefont
	{Taniguchi}, \citenamefont {Nishikawa}, \citenamefont {Yasui}, \citenamefont
	{Kobayashi}, \citenamefont {Sato}, \citenamefont {Nishioka}, \citenamefont
	{Kontani},\ and\ \citenamefont {Sano}}]{Taniguch2758}%
\BibitemOpen
\bibfield  {author} {\bibinfo {author} {\bibfnamefont {S.}~\bibnamefont
		{Taniguchi}}, \bibinfo {author} {\bibfnamefont {T.}~\bibnamefont
		{Nishikawa}}, \bibinfo {author} {\bibfnamefont {Y.}~\bibnamefont {Yasui}},
	\bibinfo {author} {\bibfnamefont {Y.}~\bibnamefont {Kobayashi}}, \bibinfo
	{author} {\bibfnamefont {M.}~\bibnamefont {Sato}}, \bibinfo {author}
	{\bibfnamefont {T.}~\bibnamefont {Nishioka}}, \bibinfo {author}
	{\bibfnamefont {M.}~\bibnamefont {Kontani}}, \ and\ \bibinfo {author}
	{\bibfnamefont {K.}~\bibnamefont {Sano}},\ }\bibfield  {title} {\enquote
	{\bibinfo {title} {Spin gap behavior of $s=1/2$ quasi-two-dimensional system
			{CaV$_4$O$_9$}},}\ }\href {\doibase 10.1143/JPSJ.64.2758} {\bibfield
	{journal} {\bibinfo  {journal} {J. Phys. Soc. Jpn.}\ }\textbf {\bibinfo
		{volume} {64}},\ \bibinfo {pages} {2758--2761} (\bibinfo {year}
	{1995})}\BibitemShut {NoStop}%
\bibitem [{\citenamefont {Moriya}(1956)}]{Moriya23}%
\BibitemOpen
\bibfield  {author} {\bibinfo {author} {\bibfnamefont {Tôru}\ \bibnamefont
		{Moriya}},\ }\bibfield  {title} {\enquote {\bibinfo {title} {Nuclear magnetic
			relaxation in antiferromagnetics},}\ }\href {\doibase 10.1143/PTP.16.23}
{\bibfield  {journal} {\bibinfo  {journal} {Progr. Theor. Phys.}\ }\textbf
	{\bibinfo {volume} {16}},\ \bibinfo {pages} {23--44} (\bibinfo {year}
	{1956})}\BibitemShut {NoStop}%
\bibitem [{\citenamefont {Gordon}\ and\ \citenamefont
	{Hoch}(1978)}]{Gordon783}%
\BibitemOpen
\bibfield  {author} {\bibinfo {author} {\bibfnamefont {M.I.}\ \bibnamefont
		{Gordon}}\ and\ \bibinfo {author} {\bibfnamefont {M.J.R.}\ \bibnamefont
		{Hoch}},\ }\bibfield  {title} {\enquote {\bibinfo {title} {Quadrupolar
			spin-lattice relaxation in solids},}\ }\href
{http://stacks.iop.org/0022-3719/11/i=4/a=023} {\bibfield  {journal}
	{\bibinfo  {journal} {J. Phys. C: Solid State Phys.}\ }\textbf {\bibinfo
		{volume} {11}},\ \bibinfo {pages} {783} (\bibinfo {year} {1978})}\BibitemShut
{NoStop}%
\bibitem [{\citenamefont {Simmons}\ \emph {et~al.}(1962)\citenamefont
	{Simmons}, \citenamefont {O'Sullivan},\ and\ \citenamefont
	{Robinson}}]{Simmons1168}%
\BibitemOpen
\bibfield  {author} {\bibinfo {author} {\bibfnamefont {W.~W.}\ \bibnamefont
		{Simmons}}, \bibinfo {author} {\bibfnamefont {W.~J.}\ \bibnamefont
		{O'Sullivan}}, \ and\ \bibinfo {author} {\bibfnamefont {W.~A.}\ \bibnamefont
		{Robinson}},\ }\bibfield  {title} {\enquote {\bibinfo {title} {Nuclear
			spin-lattice relaxation in dilute paramagnetic sapphire},}\ }\href {\doibase
	10.1103/PhysRev.127.1168} {\bibfield  {journal} {\bibinfo  {journal} {Phys.
			Rev.}\ }\textbf {\bibinfo {volume} {127}},\ \bibinfo {pages} {1168--1178}
	(\bibinfo {year} {1962})}\BibitemShut {NoStop}%
\bibitem [{\citenamefont {Nath}\ \emph {et~al.}(2009)\citenamefont {Nath},
	\citenamefont {Furukawa}, \citenamefont {Borsa}, \citenamefont {Kaul},
	\citenamefont {Baenitz}, \citenamefont {Geibel},\ and\ \citenamefont
	{Johnston}}]{Nath214430}%
\BibitemOpen
\bibfield  {author} {\bibinfo {author} {\bibfnamefont {R.}~\bibnamefont
		{Nath}}, \bibinfo {author} {\bibfnamefont {Y.}~\bibnamefont {Furukawa}},
	\bibinfo {author} {\bibfnamefont {F.}~\bibnamefont {Borsa}}, \bibinfo
	{author} {\bibfnamefont {E.~E.}\ \bibnamefont {Kaul}}, \bibinfo {author}
	{\bibfnamefont {M.}~\bibnamefont {Baenitz}}, \bibinfo {author} {\bibfnamefont
		{C.}~\bibnamefont {Geibel}}, \ and\ \bibinfo {author} {\bibfnamefont {D.~C.}\
		\bibnamefont {Johnston}},\ }\bibfield  {title} {\enquote {\bibinfo {title}
		{Single-crystal {$^{31}\text{P}$ NMR} studies of the frustrated
			square-lattice compound {Pb$_2$VO(PO$_4)_2$}},}\ }\href {\doibase
	10.1103/PhysRevB.80.214430} {\bibfield  {journal} {\bibinfo  {journal} {Phys.
			Rev. B}\ }\textbf {\bibinfo {volume} {80}},\ \bibinfo {pages} {214430}
	(\bibinfo {year} {2009})}\BibitemShut {NoStop}%
\bibitem [{\citenamefont {Ranjith}\ \emph {et~al.}(2015)\citenamefont
	{Ranjith}, \citenamefont {Majumder}, \citenamefont {Baenitz}, \citenamefont
	{Tsirlin},\ and\ \citenamefont {Nath}}]{Ranjith024422}%
\BibitemOpen
\bibfield  {author} {\bibinfo {author} {\bibfnamefont {K.~M.}\ \bibnamefont
		{Ranjith}}, \bibinfo {author} {\bibfnamefont {M.}~\bibnamefont {Majumder}},
	\bibinfo {author} {\bibfnamefont {M.}~\bibnamefont {Baenitz}}, \bibinfo
	{author} {\bibfnamefont {A.~A.}\ \bibnamefont {Tsirlin}}, \ and\ \bibinfo
	{author} {\bibfnamefont {R.}~\bibnamefont {Nath}},\ }\bibfield  {title}
{\enquote {\bibinfo {title} {Frustrated three-dimensional antiferromagnet
			{Li$_2$CuW$_2$O$_8$}: {$^{7}\mathrm{Li}$ NMR} and the effect of nonmagnetic
			dilution},}\ }\href {\doibase 10.1103/PhysRevB.92.024422} {\bibfield
	{journal} {\bibinfo  {journal} {Phys. Rev. B}\ }\textbf {\bibinfo {volume}
		{92}},\ \bibinfo {pages} {024422} (\bibinfo {year} {2015})}\BibitemShut
{NoStop}%
\bibitem [{\citenamefont {Mukhopadhyay}\ \emph {et~al.}(2012)\citenamefont
	{Mukhopadhyay}, \citenamefont {Klanj\ifmmode~\check{s}\else \v{s}\fi{}ek},
	\citenamefont {Grbi\ifmmode~\acute{c}\else \'{c}\fi{}}, \citenamefont
	{Blinder}, \citenamefont {Mayaffre}, \citenamefont {Berthier}, \citenamefont
	{Horvati\ifmmode~\acute{c}\else \'{c}\fi{}}, \citenamefont {Continentino},
	\citenamefont {Paduan-Filho}, \citenamefont {Chiari},\ and\ \citenamefont
	{Piovesana}}]{Mukhopadhyay177206}%
\BibitemOpen
\bibfield  {author} {\bibinfo {author} {\bibfnamefont {S.}~\bibnamefont
		{Mukhopadhyay}}, \bibinfo {author} {\bibfnamefont {M.}~\bibnamefont
		{Klanj\ifmmode~\check{s}\else \v{s}\fi{}ek}}, \bibinfo {author}
	{\bibfnamefont {M.~S.}\ \bibnamefont {Grbi\ifmmode~\acute{c}\else
			\'{c}\fi{}}}, \bibinfo {author} {\bibfnamefont {R.}~\bibnamefont {Blinder}},
	\bibinfo {author} {\bibfnamefont {H.}~\bibnamefont {Mayaffre}}, \bibinfo
	{author} {\bibfnamefont {C.}~\bibnamefont {Berthier}}, \bibinfo {author}
	{\bibfnamefont {M.}~\bibnamefont {Horvati\ifmmode~\acute{c}\else
			\'{c}\fi{}}}, \bibinfo {author} {\bibfnamefont {M.~A.}\ \bibnamefont
		{Continentino}}, \bibinfo {author} {\bibfnamefont {A.}~\bibnamefont
		{Paduan-Filho}}, \bibinfo {author} {\bibfnamefont {B.}~\bibnamefont
		{Chiari}}, \ and\ \bibinfo {author} {\bibfnamefont {O.}~\bibnamefont
		{Piovesana}},\ }\bibfield  {title} {\enquote {\bibinfo {title}
		{Quantum-critical spin dynamics in quasi-one-dimensional antiferromagnets},}\
}\href {\doibase 10.1103/PhysRevLett.109.177206} {\bibfield  {journal}
{\bibinfo  {journal} {Phys. Rev. Lett.}\ }\textbf {\bibinfo {volume} {109}},\
\bibinfo {pages} {177206} (\bibinfo {year} {2012})}\BibitemShut {NoStop}%
\bibitem [{\citenamefont {Xiang}\ \emph {et~al.}(2011)\citenamefont {Xiang},
	\citenamefont {Kan}, \citenamefont {Wei}, \citenamefont {Whangbo},\ and\
	\citenamefont {Gong}}]{Xiang2011}%
\BibitemOpen
\bibfield  {author} {\bibinfo {author} {\bibfnamefont {H.~J.}\ \bibnamefont
		{Xiang}}, \bibinfo {author} {\bibfnamefont {E.~J.}\ \bibnamefont {Kan}},
	\bibinfo {author} {\bibfnamefont {S.-H.}\ \bibnamefont {Wei}}, \bibinfo
	{author} {\bibfnamefont {M.-H.}\ \bibnamefont {Whangbo}}, \ and\ \bibinfo
	{author} {\bibfnamefont {X.~G.}\ \bibnamefont {Gong}},\ }\bibfield  {title}
{\enquote {\bibinfo {title} {Predicting the spin-lattice order of frustrated
			systems from first principles},}\ }\href {\doibase
	10.1103/PhysRevB.84.224429} {\bibfield  {journal} {\bibinfo  {journal} {Phys.
			Rev. B}\ }\textbf {\bibinfo {volume} {84}},\ \bibinfo {pages} {224429}
	(\bibinfo {year} {2011})}\BibitemShut {NoStop}%
\bibitem [{\citenamefont {Tsirlin}\ \emph
	{et~al.}(2011{\natexlab{b}})\citenamefont {Tsirlin}, \citenamefont {Janson},\
	and\ \citenamefont {Rosner}}]{Tsirlin2011b}%
\BibitemOpen
\bibfield  {author} {\bibinfo {author} {\bibfnamefont {A.~A.}\ \bibnamefont
		{Tsirlin}}, \bibinfo {author} {\bibfnamefont {O.}~\bibnamefont {Janson}}, \
	and\ \bibinfo {author} {\bibfnamefont {H.}~\bibnamefont {Rosner}},\
}\bibfield  {title} {\enquote {\bibinfo {title} {Unusual ferromagnetic
		superexchange in {CdVO$_3$: The role of Cd}},}\ }\href {\doibase
10.1103/PhysRevB.84.144429} {\bibfield  {journal} {\bibinfo  {journal} {Phys.
		Rev. B}\ }\textbf {\bibinfo {volume} {84}},\ \bibinfo {pages} {144429}
(\bibinfo {year} {2011}{\natexlab{b}})}\BibitemShut {NoStop}%
\bibitem [{\citenamefont {Tsirlin}\ \emph {et~al.}(2008)\citenamefont
	{Tsirlin}, \citenamefont {Nath}, \citenamefont {Geibel},\ and\ \citenamefont
	{Rosner}}]{Tsirlin2008}%
\BibitemOpen
\bibfield  {author} {\bibinfo {author} {\bibfnamefont {A.~A.}\ \bibnamefont
		{Tsirlin}}, \bibinfo {author} {\bibfnamefont {R.}~\bibnamefont {Nath}},
	\bibinfo {author} {\bibfnamefont {C.}~\bibnamefont {Geibel}}, \ and\ \bibinfo
	{author} {\bibfnamefont {H.}~\bibnamefont {Rosner}},\ }\bibfield  {title}
{\enquote {\bibinfo {title} {Magnetic properties of {Ag$_2$VOP$_2$O$_7$}: An
			unexpected spin dimer system},}\ }\href {\doibase 10.1103/PhysRevB.77.104436}
{\bibfield  {journal} {\bibinfo  {journal} {Phys. Rev. B}\ }\textbf {\bibinfo
		{volume} {77}},\ \bibinfo {pages} {104436} (\bibinfo {year}
	{2008})}\BibitemShut {NoStop}%
\bibitem [{\citenamefont {Tsirlin}\ and\ \citenamefont
	{Rosner}(2009)}]{Tsirlin2009}%
\BibitemOpen
\bibfield  {author} {\bibinfo {author} {\bibfnamefont {A.~A.}\ \bibnamefont
		{Tsirlin}}\ and\ \bibinfo {author} {\bibfnamefont {H.}~\bibnamefont
		{Rosner}},\ }\bibfield  {title} {\enquote {\bibinfo {title} {Extension of the
			spin-1/2 frustrated square lattice model: The case of layered vanadium
			phosphates},}\ }\href {\doibase 10.1103/PhysRevB.79.214417} {\bibfield
	{journal} {\bibinfo  {journal} {Phys. Rev. B}\ }\textbf {\bibinfo {volume}
		{79}},\ \bibinfo {pages} {214417} (\bibinfo {year} {2009})}\BibitemShut
{NoStop}%
\bibitem [{Note2()}]{Note2}%
\BibitemOpen
\bibinfo {note} {Similar to Ref.~\protect \rev@citealp {Tsirlin144412}, we
	perform simulations for a 2D array of alternating spin chains uniformly
	coupled by an effective interchain coupling $J_{\perp }$. The actual 3D spin
	lattice of NaVOAsO$_4$ is frustrated and thus not amenable to QMC
	simulations.}\BibitemShut {Stop}%
\end{thebibliography}
%

\end{document}